\documentclass[fleqn,usenatbib]{mnras}

\usepackage{newtxtext,newtxmath}

\usepackage[T1]{fontenc}

\DeclareRobustCommand{\VAN}[3]{#2}
\let\VANthebibliography\thebibliography
\def\thebibliography{\DeclareRobustCommand{\VAN}[3]{##3}\VANthebibliography}

\usepackage{graphicx}	
\usepackage{amsmath}	
\usepackage{hyperref}
\usepackage{natbib}
\usepackage{caption}
\usepackage{multirow}
\usepackage{geometry}
\usepackage{booktabs}
\usepackage[table]{xcolor}
\usepackage{array}
\usepackage{mathtools}
\usepackage{xspace}

\newcommand{\Reff}{\ensuremath{r_{\mathrm{e}}}\xspace}

\newcommand{\sigmaint}{\ensuremath{\sigma_{0}}\xspace}

\newcommand{\Msun}{\ensuremath{M_{\odot}}\xspace}

\newcommand{\Mstar}{\ensuremath{M_{\star}}\xspace}

\newcommand{\lMstar}{\ensuremath{\log_{10}(\Mstar/\Msun)}\xspace}

\newcommand{\logMstar}{\ensuremath{\log(\Mstar ~\rm [\Msun])}\xspace}

\newcommand{\Ha}{\ensuremath{\mathrm{H}\alpha}\xspace}
\newcommand{\vsigma}{\ensuremath{\mathrm{v}/\sigmaint}\xspace}

\newcommand{\geko}{\textsc{geko}}

\newcommand{\sersic}{S\'{e}rsic }
\newcommand{\angstrom}{\mbox{\normalfont\AA}}

\newcommand{\deltaMS}{\ensuremath{\Delta \rm MS}\xspace} 

\newcommand{\rotsupp}{\text{v}/\sigma_0} 
\newcommand{\disp}{\sigma_0}

\newcommand{\rHa}{\ensuremath{r_{\rm e, \Ha}}\xspace} 
\newcommand{\rNUV}{\ensuremath{r_{\rm e, NUV}}\xspace} 
\newcommand{\rFUV}{\ensuremath{r_{\rm e, FUV}}\xspace} 
\newcommand{\ropt}{\ensuremath{r_{\rm e, opt}}\xspace} 

\newcommand{\HatoUV}{\ensuremath{r_{\rm e, \Ha}/r_{\rm e, UV}}\xspace} 
\newcommand{\HatoNUV}{\ensuremath{r_{\rm e, \Ha}/r_{\rm e, NUV}}\xspace} 
\newcommand{\FUVtoNUV}{\ensuremath{r_{\rm e, FUV}/r_{\rm e, NUV}}\xspace} 
\newcommand{\FUVtoopt}{\ensuremath{r_{\rm e, FUV}/r_{\rm e, opt}}\xspace} 

\usepackage{adjustbox}  
\newcommand\orcid[1]{\href{http://orcid.org/#1}{\adjustbox{trim={-.15\width} {0\height} {-.15\width} {0\height},clip}{\includegraphics[height=10pt]{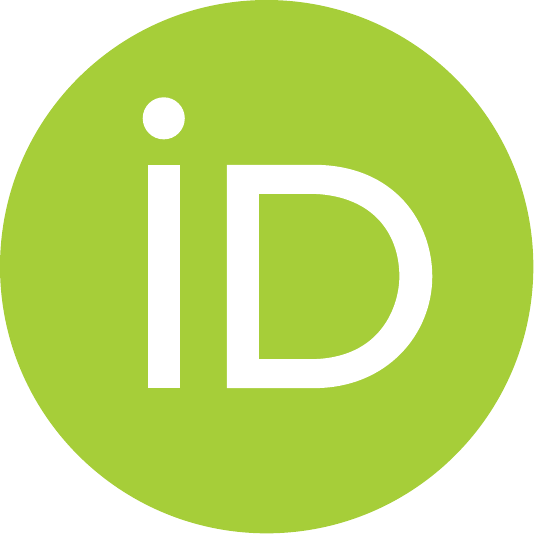}}}}

\title[Multi-wavelength size-mass relations at z>4]{Beyond the stars: Linking H$\alpha$ sizes, kinematics, and star formation in galaxies at $z\approx 4-6$ with \textit{JWST} grism surveys and \textsc{geko}}

\author[A. L. Danhaive et al.]{A. Lola Danhaive\orcid{0000-0002-9708-9958}$^{1,2}$\thanks{ald66@cam.ac.uk},
Sandro Tacchella\orcid{0000-0002-8224-4505}$^{1,2}$, William McClymont\orcid{0009-0009-5565-3790}$^{1,2}$, Brant Robertson\orcid{0000-0002-4271-0364}$^{3}$,  \newauthor Stefano Carniani\orcid{0000-0002-6719-380X}$^{4}$, Courtney Carreira\orcid{0000-0001-6301-3667}$^{3}$, Eiichi Egami\orcid{0000-0003-1344-9475}$^{5}$, Andrew J.\ Bunker \orcid{0000-0002-8651-9879}$^{6}$,
\newauthor Emma Curtis-Lake\orcid{0000-0002-9551-0534}$^{7}$,  Daniel J.\ Eisenstein\orcid{0000-0002-2929-3121}$^{8}$, Zhiyuan Ji\orcid{0000-0001-7673-2257}$^{5}$, Benjamin D. Johnson\orcid{0000-0002-9280-7594}$^{8}$,  Marcia Rieke\orcid{0000-0002-7893-6170}$^{5}$, 
\newauthor Natalia C. Villanueva\orcid{0000-0001-6917-4656}$^{1,2,9}$, Christopher N. A. Willmer\orcid{0000-0001-9262-9997}$^{5}$, Chris Willot\orcid{0000-0002-4201-7367}$^{10}$,  Zihao Wu\orcid{0000-0002-8876-5248}$^{8}$, Yongda Zhu\orcid{0000-0003-3307-7525}$^{5}$
\\
\\
$^{1}$Kavli Institute for Cosmology, University of Cambridge, Madingley Road, Cambridge, CB3 0HA, UK \\
$^{2}$Cavendish Laboratory, University of Cambridge, 19 JJ Thomson Avenue, Cambridge, CB3 0HE, UK\\
$^{3}$ Department of Astronomy and Astrophysics, University of California, Santa Cruz, 1156 High Street, Santa Cruz, CA 95064, USA \\
$^{4}$ Scuola Normale Superiore, Piazza dei Cavalieri 7, I-56126 Pisa, Italy \\
$^{5}$ Steward Observatory, University of Arizona, 933 N. Cherry Avenue, Tucson, AZ 85721, USA \\
$^{6}$ Department of Physics, University of Oxford, Denys Wilkinson Building, Keble Road, Oxford OX1 3RH, UK \\
$^{7}$ Centre for Astrophysics Research, Department of Physics, Astronomy and Mathematics, University of Hertfordshire, Hatfield AL10 9AB, UK \\
$^{8}$ Center for Astrophysics $|$ Harvard \& Smithsonian, 60 Garden St., Cambridge MA 02138 USA \\
$^{9}$The University of Texas at Austin, Department of Astronomy, 2515 Speedway, Stop C1400, Austin, TX 78712-1205, USA \\
$^{10}$ NRC Herzberg, 5071 West Saanich Rd, Victoria, BC V9E 2E7, Canada \\
}

\date{Accepted XXX. Received YYY; in original form ZZZ}

\pubyear{2015}

\begin{document}
\label{firstpage}
\pagerange{\pageref{firstpage}--\pageref{lastpage}}
\maketitle

\begin{abstract}
Understanding how galaxies assemble their mass during the first billion years of cosmic time is a central goal of extragalactic astrophysics, yet joint constraints on their sizes and kinematics remain scarce. We present one of the first statistical studies of the \Ha\ size-mass relation at high redshift with a sample of 213 galaxies at spectroscopic redshifts of $z\approx 4-6$ from the FRESCO and CONGRESS NIRCam grism surveys. We measure the \Ha\ morphology and kinematics of our sample using the novel forward modelling Bayesian inference tool \geko, and complement them with stellar continuum sizes in the rest-frame FUV, NUV, and optical, obtained from modelling of imaging data from the JADES survey with \textsc{Pysersic}. At $z\approx5$, we find that the average H$\alpha$ sizes are larger than the stellar continuum (FUV, NUV and optical), with $\rHa = 1.17 \pm 0.05$ kpc and $r_{\rm e,cont} \approx 0.9$ kpc for galaxies with $\logMstar = 9.5$. However, we find no significant differences between the stellar continuum sizes at different wavelengths, suggesting that galaxies are not yet steadily growing inside-out at these epochs. Instead, we find that the ratio $\HatoNUV$ increases with the distance above the star-forming main sequence ($\deltaMS$), consistent with an expansion of H$\alpha$ sizes during episodes of enhanced star formation caused by an increase in ionising photons. As galaxies move above the star-forming main sequence, we find an increase of their rotational support $\rotsupp$, which could be tracing accreting gas illuminated by the \Ha\ emission. Finally, we find that about half of the elongated systems ($b/a < 0.5$) are not rotationally supported, indicating a potential flattened/prolate galaxy population at high redshift. 
\end{abstract}




\section{Introduction}

Galaxy morphology is a powerful probe of galaxy evolution over cosmic time. As galaxies build up their stellar masses through different physical processes, the spatial evolution of different components traces how star formation, chemical enrichment, and dust content evolves in galaxies. This interplay is best illustrated by the relation between galaxy sizes, often parametrized through their half-light radius $\Reff$, and their stellar mass $\Mstar$, known as the size-mass relation \citep[SMR;][]{Shen:2003aa,Kauffmann:2003aa}. Beyond the sizes of galaxies, this relation traces complex processes such as mergers \citep{Robertson:2006aa,Naab:2009aa, Hopkins:2009aa,Oser:2010aa,Wellons:2015aa}, feedback from star formation and active galactic nuclei \citep{Sijacki:2007aa,Hopkins:2014aa,Agertz:2015aa,Dubois:2016aa}, and radial dust attenuation profiles \citep{Nelson:2016aa,Tacchella:2018wf, Matharu:2023uj}. This is exemplified through the redshift evolution of the size-mass relation with redshift \citep{van-der-Wel:2014ab,Mowla:2019ab,Suess:2019aa,Mosleh:2020aa}, and the large scatter found at fixed redshift, which has been shown to depend strongly on whether galaxies are star-forming or quiescent \citep{Carollo:2013aa,Fagioli:2016aa,Nedkova:2021aa}.

The sizes of galaxies at different wavelengths probe the spatial extent of different stellar populations. Nebular lines such as $\Ha$ trace the ionized gas of galaxies and are closely related to emission from young ($\lesssim$10 Myr) massive stars and hence ongoing star formation. The UV continuum traces young stars directly, but also has contributions from older stellar populations ($\sim$10-100 Myr). On the other hand, optical continuum traces even older stellar populations. More importantly, it typically traces the bulk of the stellar mass, making it one of the most common probes of galaxy sizes. However, these considerations do not necessarily hold at high redshift $z>3$ due to strong outshining effects \citep{Tacchella:2022tn,Papovich:2023aa,Tacchella:2023aa, Whitler:2023aa}. This causes the continuum from UV to optical to be dominated by emission from young stars, meaning that they all end up tracing similar timescales of star formation.

In analytical models of galaxy formation, gas accretes a dark matter (DM) halo, then cools to form stars. As such, the sizes of galaxies are directly related to the mass of their host halos, $R \propto M_{\rm halo}^{1/3}$ \citep{Fall:1980aa,Mo:1998aa,Kravtsov:2013aa,Somerville:2018aa}, where $M_{\rm halo}$ is the mass of the DM halo and $R$ is a characteristic radius such as the virial radius. Because of this fundamental relation, it is also expected that the sizes of galaxies, $\Reff$, are directly linked to their stellar mass through a similar power-law relation, $\Reff\propto M_{\star}^\alpha$ \citep{van-der-Wel:2014ab,Mowla:2019ab,Nedkova:2021aa}. This framework motivated the parametrization of the SMR as 
\begin{equation}
    \log \Reff = \alpha \log(M_{\star}/M_0) + \beta,
\end{equation}
\noindent
where $M_0$ is a mass normalisation factor, and $\beta$ is the normalization of the SMR at $\Mstar = M_0$. These parameters can be measured across wavelengths and redshifts, their evolution containing information on the physical processes that drive galaxy sizes, and hence galaxy growth, at different times. Following this same reasoning, we expect galaxies to grow "inside-out" with the average $\Reff$ for a given stellar mass increasing with cosmic time. This directly follows from the growth of the virial radii of dark-matter haloes, $R_{\rm vir}\propto (1+z)^{-1}$, driven by the expansion of the Universe \citep{Fall:1980aa,Mo:1998aa,Ferguson:2004aa}.

\textit{The Hubble Space Telescope (HST)} probed the rest-frame optical continuum out to $z\approx3$ and the rest-frame UV out to $z\approx 10$, paving the way for the first measurements of the SMR across large redshift ranges. These studies found that the optical SMR of star-forming galaxies at cosmic noon ($z\approx 1-3$) has a positive slope $\alpha>0$ \citep{van-der-Wel:2014ab, Shibuya:2015aa,Mowla:2019aa}, in agreement with measurements in the local Universe \citep{Shen:2003aa}. However, the evolution of this slope with redshift and mass is still contested. Key differences became clear between star-forming and quiescent galaxies. When star-forming galaxies undergo quenching, their growth mode changes from \textit{in-situ} star formation in an extended gas disk to growth preferentially through dry mergers \citep{Robertson:2006aa, Naab:2009aa, Oser:2010aa}. This causes a steepening of the SMR for quiescent galaxies \citep{van-Dokkum:2015aa}. Furthermore, spatially resolved studies of nebular emission (through \Ha) and stellar continuum \citep[e.g.][]{Forster-Schreiber:2011aa} found evidence for inside-out growth through extended \Ha\ sizes and equivalent widths (EWs) \citep{Nelson:2016wo}. However, this growth is weak, consistent with radially constant specific star-formation rate (sSFR) profiles and a shallow slope of the SMR \citep{Tacchella:2015aa,Tacchella:2018wf,Morselli:2019aa}, and studies have postulated that the observed SMR is mainly driven by effects of the mass-to-light ratio \citep{Szomoru:2013aa,Mosleh:2017aa,Suess:2019aa}.

With the arrival of the \textit{James Webb Space Telescope (\textit{JWST})}, rest-frame optical sizes have been measured beyond $z\approx3$ for star-forming galaxies \citep{Ward:2024aa,Miller:2024aa, Varadaraj:2024aa,Matharu:2024aa,Yang:2025aa,Allen:2025aa} and for quiescent galaxies \citep{Ji:2024aa,Ito:2024aa,Yang:2025aa}, and show good agreement with the rest-UV out to $z\approx 6$. Interestingly, \citet{Allen:2025aa} find that the size scatter is larger in the rest-UV compared to the rest-optical, which could be evidence for bursty star formation, which is expected at high redshift \citep{Tacchella:2016ab,Boyett:2024ab,McClymont:2025aa,Simmonds:2025aa,Witten:2025aa}. Such studies highlight the link between morphology and star formation, making sizes an important tracer of galaxy growth across cosmic time. However, the UV continuum is also much more sensitive to dust attenuation than the optical, making the direct comparison more difficult. 

At fixed stellar mass, the size of galaxies increases with cosmic time \citep{van-der-Wel:2014ab, Allen:2025aa,Ward:2024aa}, suggesting a more regular build-up of a star-forming disk around a central bulge, at later cosmic times. Although at higher redshift we typically probe lower masses $\logMstar\leq 10$, colour gradients can be found that potentially indicate the presence of spatially distinct old and young stellar populations \citep{Baker:2025ab}. Because of the growth of structure in the Universe, dictated by the scale factor $(1+z)$, we expect this increase in sizes with cosmic time, often parametrized as $\Reff\propto(1+z)^{\gamma}$. Similar to the slope of the SMR, the value and evolution of the growth factor $\gamma$ is contested \citep[e.g. see][and references therein]{Shibuya:2015aa}.

The sizes, and more generally the morphologies, of galaxies are also linked with their kinematic state. In the local universe, spiral galaxies have been associated with cold rotating disks, whereas spheroids often show less rotation and more pressure support \citep{Kormendy:2004aa, Simard:2011aa}. Across redshifts, mergers are associated with disturbed morphologies and kinematics \citep{Wright:2009ub,Qu:2010aa,Duan:2025aa,Puskas:2025aa}. However, it is not always straightforward to infer the kinematic state of galaxies from their morphologies. For example, studies have shown that elliptical galaxies have a range of rotational support \citep{Emsellem:2007aa, Cappellari:2016aa}, and that resolution limits, driven by the point-spread function (PSF), can cause observations to appear disk-like in shape (with \sersic indices $n\approx 1$) while concealing merging or clumpy systems \citep{Simons:2019aa}. It is hence important to push morpho-kinematic studies of galaxies to high-redshift \citep[e.g.][]{Wisnioski:2015vx,Wisnioski:2019tg, Jones:2021aa,Jones:2024ab, Marconcini:2024ab,Parlanti:2025aa}. 

Emission lines, which are often used to probe gas kinematics, are an ideal laboratory to directly compare galaxy morphology and dynamics. At cosmic noon ($1<z<3$), $\Ha$ sizes show a large scatter that has a weak dependence on redshift and stellar mass \citep{Wilman:2020aa}. With \textit{JWST}, studies of $\Ha$ maps have gone from cosmic noon \citep[][]{Wisnioski:2015vx,Nelson:2016wo,Forster-Schreiber:2018aa,Genzel:2020aa} to high redshift \citep[$z>4$;][]{Nelson:2024aa, Matharu:2024aa, Danhaive:2025aa,Stephenson:2025aa}. Using scaling relations \citep{Kennicutt:1998vu} and/or spectral energy distribution (SED) fitting, $\Ha$ maps can be used to infer star-formation rate (SFR) profiles, which directly map the spatially resolved growth of galaxies \citep{Tacchella:2015aa,Nelson:2016wo,Belfiore:2018aa}. However, this mapping relies on the assumption that the emitted Lyman continuum (LyC) photons are absorbed locally, which is valid in massive star-forming galaxies at lower-redshift, where most LyC photons are absorbed on scales significantly smaller than the effective radius \citep[$\lesssim$1\,kpc;][]{McClymont:2024aa}. However, this assumption breaks down at low masses and high redshifts where the galaxy sizes become comparable to the mean free path of LyC photons \citep{McClymont:2025ab}. The detection of emission lines also offers an important constraint on the redshift of observed galaxies, which is lacking in many photometry-only based studies. When trying to quantify the redshift evolution of the SMR, benefitting from spectroscopic redshifts is crucial.

The Near Infrared Camera \citep[NIRCam; ][]{Rieke:2023va} aboard \textit{JWST} is ideally suited to combine morphology and kinematics at high redshift. It offers high resolution ($\approx 0.03-0.15''$) imaging at $0.6-5 ~\mu$m, probing stellar continuum from the far ultraviolet (FUV, $\lambda \approx 1500 ~\angstrom$) to the optical ($\lambda \approx 5500~\angstrom$). Its grism mode provides wide-field slitless spectroscopy (WFSS) for the entire NIRCam field-of-view, providing 2D spectra of stellar continuum and emission lines with medium spectral resolution $R = \lambda/\Delta\lambda \approx 1600$ ($\sigma = (1/2.36)\times c/R= 80-90$ km/s at $\lambda = 3-5 \thinspace \mu$m). The combination of NIRCam imaging and grism data opens the door for large statistical studies of stars and ionized gas out to $z\approx 9$. Importantly, the 2D grism spectra are encoded with kinematic information, which can be modelled and recovered using Bayesian inference. This framework is the basis of the Grism Emission-line Kinematics tOol (\textsc{geko}; Danhaive et al. in prep), an inference forward-modelling tool which enabled the first statistical study of ionized gas kinematics beyond cosmic noon \citep{Danhaive:2025aa}.

In this work, we present one of the first $\Ha$ SMRs at $z>4$, obtained from morpho-kinematic modelling of \textit{JWST}/NIRCam grism data from the First Reionization Epoch Spectroscopically Complete Observations survey \citep[FRESCO;][PI: Oesch, PID: 1895]{Oesch:2023aa}, in both GOODS fields \citep{Giavalisco:2004aa}, and the COmplete Nircam Grism REdShift Survey (CONGRESS, PIs: Egami, Sun, PID: 3577), in the GOODS-N field. In \cite{Danhaive:2025aa}, we use \geko\ to combine NIRCam grism measurements with NIRCam multi-band imaging from the \textit{JWST} Advanced Deep Extragalactic Survey \citep[JADES;][PIs: Rieke and Curtis-Lake]{Bunker:2020aa,Rieke:2020aa,Eisenstein:2023aa} in order to infer kinematic constraints on $\approx 200$ \Ha\ emitters at $z\approx4-6$. In this paper, we focus specifically on the \Ha\ sizes and the morphologies, combined with measurements of sizes in the rest-frame near-UV (NUV), FUV, and optical obtained from modelling images from the JADES survey. We combine this multi-wavelength information with measurements of the galaxy kinematics and star formation properties to study galaxy growth at $z>4$. 

This paper is organized as follows. We present our sample in Sec. \ref{sec:data}, which is derived from \citet{Danhaive:2025aa}, describing both the imaging and spectroscopy components. We also outline the morphological modelling techniques used to derive size measurements for the \Ha\ and the multi-wavelength stellar continuum. We present our results for the SMRs at $z\approx 4-6$ in Sec. \ref{sec:res-SMR}, placing them in the context of other published studies covering a range of redshifts. In Sec. \ref{sec:mutliwave}, we explore how the evolution of different size ratios with star formation and stellar mass can shed light on the evolution of galaxies along the main sequence. We link our morphological measurements with kinematics in Sec. \ref{sec:growth-and-kins}. We interpret our findings in Sec. \ref{sec:discussion} to understand what physical processes drive the observed growth of galaxies at high redshift. Finally, we summarize our findings in Sec. \ref{sec:conclusions}. Throughout this work we assume $\Omega_0 =0.315$ and $H_0 = 67.4 \thinspace \text{km}\thinspace\text{s}^{-1}\thinspace\text{Mpc}^{-1}$ \citep{Planck-Collaboration:2020aa}.

\section{Observations, sample and methodology}\label{sec:data}
In this section, we present the data used in this work (Secs. \ref{sec:grism-obs} and \ref{sec:data-imaging}), outlining its reduction and the sample selection criteria (Sec. \ref{sec:sample-selection}). Our sample is based on the complete sample from \citet{Danhaive:2025aa}, where a more detailed description can be found. In order to measure the sizes from the \Ha\ emission and the stellar continuum, we need to model their morphology, accounting for instrumental effects. We detail our methodology for the NIRCam imaging data in Sec. \ref{sec:pysersic-mod} and the NIRCam grism data in Sec. \ref{sec:morph-kin-mod}.

\subsection{NIRCam grism}\label{sec:grism-obs}

For the analysis of the ionized gas morphology and kinematics, we use NIRCam grism data in F444W filter ($3.9-5.0 \thinspace\mu$m) obtained with FRESCO \citep{Oesch:2023aa} in both GOODS fields and in the F356W filter ($3-4 \thinspace\mu$m) obtained with CONGRESS (PIs: Egami, Sun, PID: 3577) in GOODS-N. These filters are chosen as they probe the \Ha\ emission ($\lambda_{\rm rest} = 6565 ~\angstrom$) at $z=3.8-6.5$. The observations for both surveys are taken using the row-direction grisms (GRISMR mode) with modules A and B, with a spectral resolution of $R\approx 1400-1650$ across the filters used. The reduction technique used for the grism data is detailed in \citet{Sun:2023ab} and \citet{Helton:2024aa}. The continuum subtracted emission-line maps used for the morpho-kinematic modelling (Sec. \ref{sec:morph-kin-mod}) are obtained by applying row-by-row median filtering using a 2-step iterative technique as described in \citet{Kashino:2023wv}.

\subsection{NIRCam imaging} \label{sec:data-imaging}

The FRESCO and CONGRESS grism programs also obtained imaging data in the F182M and F210M, and F090W and F115W filters, respectively. In addition, there is a wealth of imaging data in the GOODS fields from the JADES survey \citep{Eisenstein:2023aa,Rieke:2023aa}, a Guaranteed Time Observations (GTO) programme of the NIRCam and NIRSpec instrument teams that obtained imaging of an area of $\approx 175$ arcmin$^2$ with 8-10 filters. This overlap provides additional imaging in the wide bands F090W, F115W, F150W, F200W, F277W, F356W, and deeper F444W observations, and medium bands F335M and F410M observations. For certain regions in our sample, we also have additional medium bands F182M, F210M, F430M, F460M, and F480M from the \textit{JWST}\textit{ }Extragalactic Medium Band Survey \citep[JEMS;][]{Williams:2023aa} in GOODS-S. This wide range of filters is not only optimal to probe sizes across different wavelengths, but also to provide tighter constraints on the SED modelling (Sec. \ref{sec:sed-mod}). We use high-resolution data and photometric catalogues obtained from drizzled mosaics, and the full details on the data reduction, catalogue generation, and photometry can be found in \citet{Rieke:2023aa,Robertson:2024aa}. 

\subsection{Sample selection}\label{sec:sample-selection}
Our sample is taken from \citet{Danhaive:2025aa}, where a full description of the sample selection can be found, but we offer a summary here. The sample of \Ha\ emitters from the FRESCO and CONGRESS surveys is built using photometric redshift estimates from \textsc{EAZY} \citep{Brammer:2008aa,Hainline:2024aa}, which are then verified through Gaussian modelling of the 1D integrated spectrum containing the \Ha\ line, amongst others. The derived spectroscopic redshifts are then confirmed through visual inspection. This process resulted in a catalogue of $>1000$ \Ha\ emitters with S/N>5 detections, which is in good agreement with previous work using the same grism data \citep{Covelo-Paz:2025aa}. In this work, we make a more conservative cut of $\rm S/N > 10$,  and obtain 393 galaxies in CONGRESS and 189 in FRESCO \citep{Helton:2024aa,Lin:2025aa}.  

However, in order to recover reliable constraints on the kinematics from grism data, we need not only high enough S/N, but also a favourable galaxy orientation on the sky. In fact, if the major axis is aligned to the grism dispersion direction, the effects of the velocity and velocity dispersion are completely degenerate with each other, and with the size and morphological profile of the galaxy. We therefore require that the position angle (PA) of the galaxy, defined from the positive y-axis, be distinct from the dispersion direction (defined by $\rm PA = 90^{\circ}$). To achieve this, we only select galaxies with $\rm PA < 75^{\circ}$. We also remove objects that have multiple components and therefore are not well modelled by a single \sersic profile, such as merging systems. The PA and S/N cuts can be more or less conservative, resulting in the three sub-samples presented in \citet{Danhaive:2025aa}:

\begin{itemize}
    \item \textit{Gold sample:} This sample has the most conservative selection cut, and hence contains galaxies for which we have the most reliable constraints. We enforce a high integrated \Ha\ S/N ($\mathrm{S/N}>20$) and have a PA satisfying $|\text{PA}_{\text{morph}}| < 60^{\circ}$. This sample contains 37 galaxies. 
    
    \item \textit{Silver sample:} This less conservative sample contains 126 galaxies. It only requires $\text{S/N}\geq 10$ and has a looser PA cut $|\text{PA}_{\text{morph}}| < 75^{\circ}$.
    
    \item \textit{Unresolved sample: }This sample contains $\text{S/N}\geq 10$ galaxies whose inferred \Reff\ is smaller than half of the full-width-half-maximum (FWHM) of the F444W PSF and/or whose velocity gradient is unresolved, i.e., its value is consistent with zero within $1\sigma$: $\rm|v(r=r_{\rm e})|/\Delta v(r=r_{\rm e}) < 1$, where $\Delta v(r=r_{\rm e})$ is the 1$\sigma$ lower limit. This sample contains 50 galaxies.

\end{itemize}

In this paper we use the full sample of 213 galaxies in our analyses and make no further distinction between the three subsamples. This distinction was important in \cite{Danhaive:2025aa} where the focus is on kinematics, but in this morphology-centred paper we prioritize sample size over tight kinematic constraints. 

\subsection{SED modelling}\label{sec:sed-mod}

As detailed in \citet{Danhaive:2025aa}, we infer stellar masses, star formation rates averaged over different timescales, and constraints on the dust attenuation from SED modelling with \textsc{Prospector} \citep{Johnson:2021aa}. We fit all of the available photometry from JADES (Sec. \ref{sec:data-imaging}), including MIRI imaging in F770W and F1280W, in conjunction with additional bands from JEMS and FRESCO  (F182M, F210M, F430M, F460M, and F480M), and available photometry from \textit{HST}'s Advanced Camera for Surveys (ACS) bands F435W, F606W, F775W, F814W, and F850LP \citep[the GOODS survey][]{Dickinson:2004aa,Giavalisco:2004aa} released through the Hubble Legacy Field program \citep{Whitaker:2019aa}. Following \citet{Tacchella:2023aa}, we model the photometry self-consistently with the available \Ha\ line fluxes from the FRESCO and CONGRESS surveys, and, for FRESCO galaxies in the GOODS-N field, we also include, when detected, $[\text{OIII}]5007$\AA, $[\text{OIII}]4959$\AA, H$\beta$, and [SIII]$9533$\AA\ fluxes from the CONGRESS survey. The redshift in the SED fitting is fixed to the spectroscopic redshifts derived from the grism spectra, which allows us to break the degeneracy between redshift and physical properties such as stellar mass. 

\begin{figure}
    \centering
    \includegraphics[width=1\linewidth]{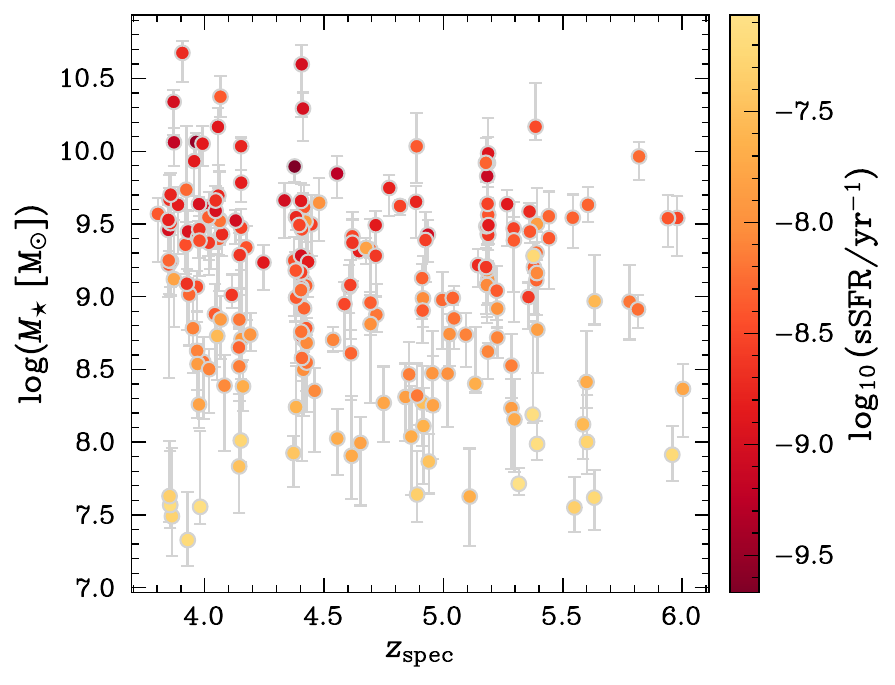}
    \caption{Stellar mass ($M_{\star}$) as a function of spectroscopic redshift ($z_{\rm spec}$) for our sample of 213 galaxies, colour-coded by sSFR, average over 10 Myr. As expected, we are biased to high sSFR systems at low stellar masses, as these systems will be the brightest in \Ha\ and pass our S/N cut. }
    \label{fig:mass-z}
\end{figure}

We present our sample in Fig. \ref{fig:mass-z}, where we show the stellar mass as a function of redshift. We colour-code our sample by specific star-formation rate (sSFR), which is defined as $\rm sSFR = SFR/\Mstar$ and is a good probe of where galaxies lie with respect to the star-forming main sequence (SFMS). It is apparent that, at low masses $\logMstar\leq8.5-9$, we are biased to highly star-forming systems with high sSFRs. This is a direct result of our integrated \Ha\ flux S/N cut. The sSFR measures the SFR per unit stellar mass of a galaxy, and it roughly equals the inverse of the mass doubling timescale $t_{\rm double} = 1/\rm sSFR$. This implies that the galaxies in our sample, if they keep a constant sSFR, would double their stellar masses in only  $\lesssim 30 ~\rm Myr$. Although galaxies oscillate around the SFMS and do not have constant sSFRs, this still highlights the short timescales on which galaxies grow at $z\approx4-6$.

\subsection{Morphological modelling with \texttt{pysersic}} \label{sec:pysersic-mod}

In order to constrain the sizes of the galaxies in our sample at different rest-frame wavelengths, we use the array of available wide band imaging from the JADES survey. For galaxies at $z=3.8-5.0$ ($z=5.0-6.0$), the rest-frame FUV ($\lambda\approx 1500$ \angstrom), NUV ($\lambda\approx 2800$ \angstrom), and optical ($\lambda\approx 5500$ \angstrom) continuum are well-probed by the F090W, F150W, and F277W (F090W, F200W, and F356W) filters (Tab. \ref{tab:filters}). The imaging in all of these filters has a comparable exposure time ($\sim$ 5h) for most of the galaxies in our sample. The $\Ha$ sizes are constrained using morpho-kinematic modelling of the F356W and F444W grism data, respectively (see Sec. \ref{sec:morph-kin-mod}).

\renewcommand{\arraystretch}{2}
\begin{table}
\centering
\begin{tabular}{c|c|c|c|c}
Redshift & FUV & NUV & Optical & $\Ha$ \\ \hline
     $z=3.8-5.0$    &     F090W   & F150W & F277W &  F356W  \\ \hline
      $z=5.0-6.0$   &    F090W    &    F200W     & F356W& F444W  
\end{tabular}
\caption{NIRCam filters used to trace emission at FUV, NUV, optical, and $\mathrm{H}\alpha$ wavelengths as a function of redshift, which are modelled to measure galaxy sizes.}
\label{tab:filters}
\end{table}

We fit the relevant wide band images with the fully Bayesian code \textsc{pysersic} \citep{Pasha:2023aa}. \textsc{pysersic} models imaging data from any filter to infer the best-fit \sersic profile(s) parameters. \textsc{pysersic} forward-models the image and accounts for the PSF convolution, so that the resulting morphological parameter are intrinsic, meaning PSF-deconvolved. We approximate the PSF in each filter using the model PSFs (mPSFs) constructed by mosaicking WebbPSF models repeatedly over the field identically to our exposure mosaics and then measuring the average PSF. This method is presented in \citet{Ji:2024aa}. The model PSFs are then resample down to the specific filter pixel resolution and used for modelling. In terms of sampling, we use the stochastic variational inference (SVI)-flow mode of \textsc{pysersic}, where the posterior distributions of the model parameters are inferred using normalizing flows. We note that this method is consistent with results for the same objects obtained using the \textsc{pysersic} Markov Chain Monte Carlo (MCMC) sampling mode (Carreira et al., in prep.).

\begin{figure}
    \centering
    \includegraphics[width=1\linewidth]{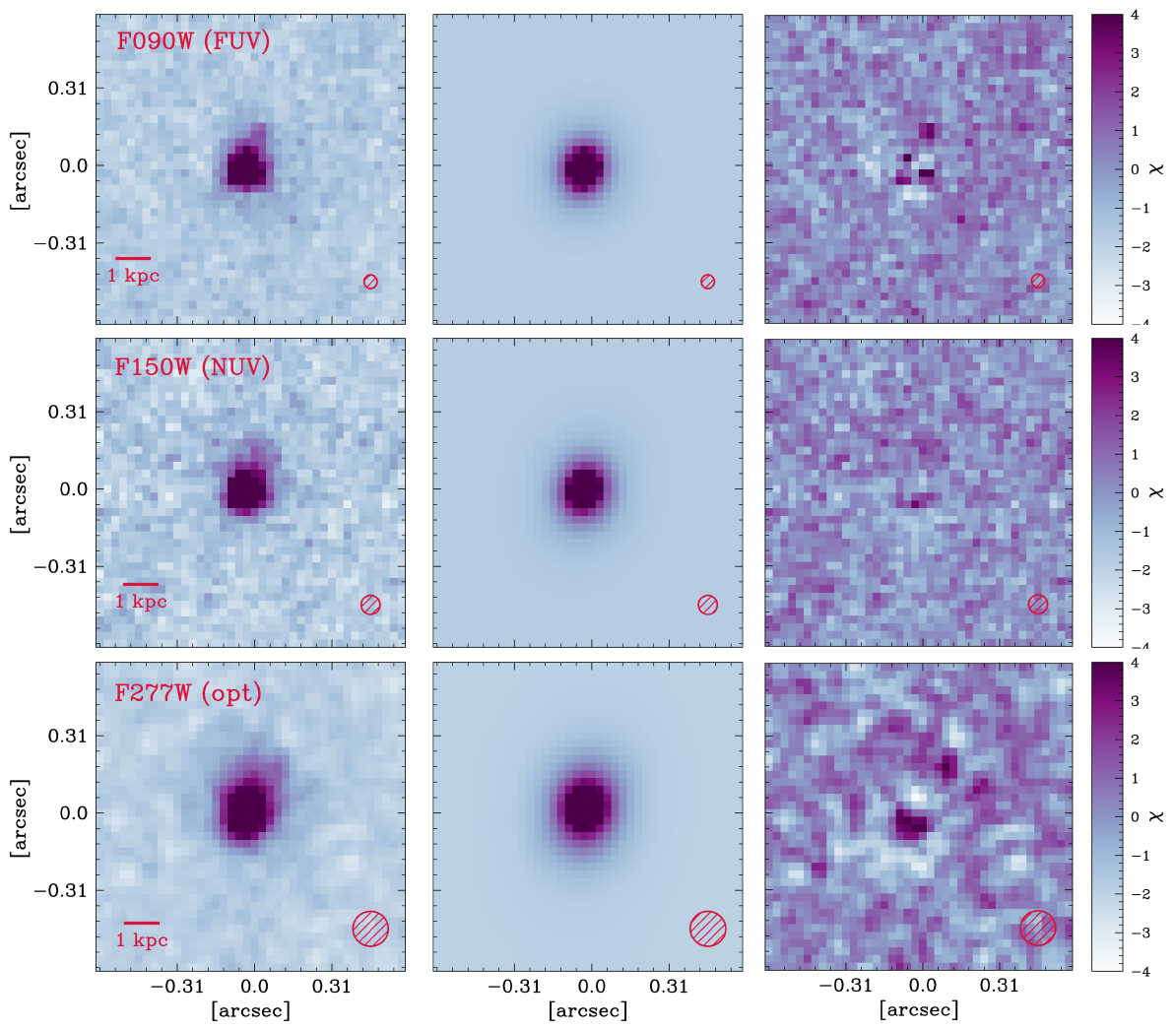}
    \caption{From top to bottom: rest-frame FUV, NUV and optical \textsc{pysersic} fits for a galaxy in our sample, probed by the NIRCam filters F090W, F200W, and F356W at $z\approx5.2$. In each band, the galaxy (left) is modelled with a PSF (red circle)-convolved one-component \sersic profile (middle), and contaminants are masked during the fitting. The residual images (right) show signs of a clumpy morphology which is not fully captured by this simple model. However, the overall size (effective radius) of the galaxy is well measured by the model.}
    \label{fig:py-fits}
\end{figure}

We model our galaxies with a one-component \sersic profile \citep{Sersic:1968aa}:

\begin{equation}
    I(r) = I_e \exp\Biggl( -b_n\left[ \left(\frac{r}{r_{\rm e}}\right)^{1/n}-1\right]\Biggr),
\label{eq:Sersic_profile}
\end{equation}
\noindent
and derive measurements of position angle $\theta$, ellipticity $e$, \sersic index $n$, light centroid $(x_0,y_0)$, and, importantly, half-light radius $r_{\rm e}$. Each band is independently fit. Additionally, we fit for a flat sky background and we mask any contaminant source from the cutout. Fig. \ref{fig:py-fits} shows the fits for a $z\approx 5.2$ galaxy in our sample, as an example for our morphological modelling procedure. Despite some underlying structure present in the residuals, we obtain a satisfactory fit given our simple model. Specifically, this structure is not expected to strongly affect the measurement of sizes, which is dominated by the main bright component at high-redshift \citep[e.g.][]{Mosleh:2013aa}.

\subsection{Morpho-kinematic modelling with \texttt{geko}}\label{sec:morph-kin-mod}

Using the \textsc{geko} code, we forward-model specific emission line, in this case $\Ha$, morphology and kinematics and fit the NIRCam grism data using a Bayesian inference framework. The full methodology is detailed in \citet{Danhaive:2025aa} and the full code will be presented in an upcoming release paper (Danhaive et al., in prep), but we summarize the key elements here. 

The emission line morphology is modelled with a single \sersic profile with the six free parameters outlined in Sec. \ref{sec:pysersic-mod}. To generate a prior for these parameters, needed to help alleviate the morphology-kinematics degeneracy of the 2D grism data, we use results from our \textsc{pysersic} modelling of the F150W band, which probes the UV continuum in both redshift ranges. For galaxies not covered by F150W, we use the nearby F182M filter. The width of the priors are estimated by doubling the uncertainties derived with \textsc{pysersic} to account for the fact that we are not measuring the true emission-line morphology, but stellar continuum emission which probes similar star-formation timescales as \Ha.

The morphological model is then convolved with the kinematics before being convolved with the instrument's PSF and line-spread function (LSF) and projected in the grism 2D space. The kinematics are modelled with an arctangent velocity curve \citep{Courteau:1997uw, Miller:2011aa}
\begin{equation}
    V_{\text{rot}}(r_{\text{int}}, r_{\text{t}}, V_a) = \frac{2}{\pi}V_a\arctan{\frac{r_{\text{int}}}{r_{\text{t}}}},
\end{equation}
and a constant velocity dispersion. The inclination of the velocity curve is set to the value derived from the morphological fit. The free kinematic parameters are hence: $V_a$, the asymptotic velocity, $r_t$, the turn-around radius, $\rm PA_{kin}$, the position angle which is fit independently from the morphological one, and intrinsic velocity dispersion $\sigma_0$.  The kinematic parameters have uniform priors, with $r_t$ forced to be below $r_{\rm e}$, motivated by lower redshift studies finding $r_t \approx 0.25 \thinspace r_{\rm e}$ \citep{Miller:2011aa}. The prior for the kinematic position angle $\rm PA_{kin}$ is the same as for $\rm PA_{morph}$, despite $\rm PA_{kin}$ being fit independently.

\section{The size-mass relation across cosmic time}\label{sec:res-SMR}
In this section, we present our analysis of the SMR at different rest-frame wavelengths (FUV, NUV, optical, and \Ha) at $z\approx 4-6$, and place our results in the context of other studies across cosmic time. We discuss the normalization (Sec. \ref{sec:size-evol}), the mass dependence (Sec. \ref{sec:slope-evol}), the redshift dependence (Sec. \ref{sec:z-evol}), and the intrinsic scatter (Sec. \ref{sec:scatter-evol}).
\begin{figure*}
    \centering
    \includegraphics[width=1\linewidth]{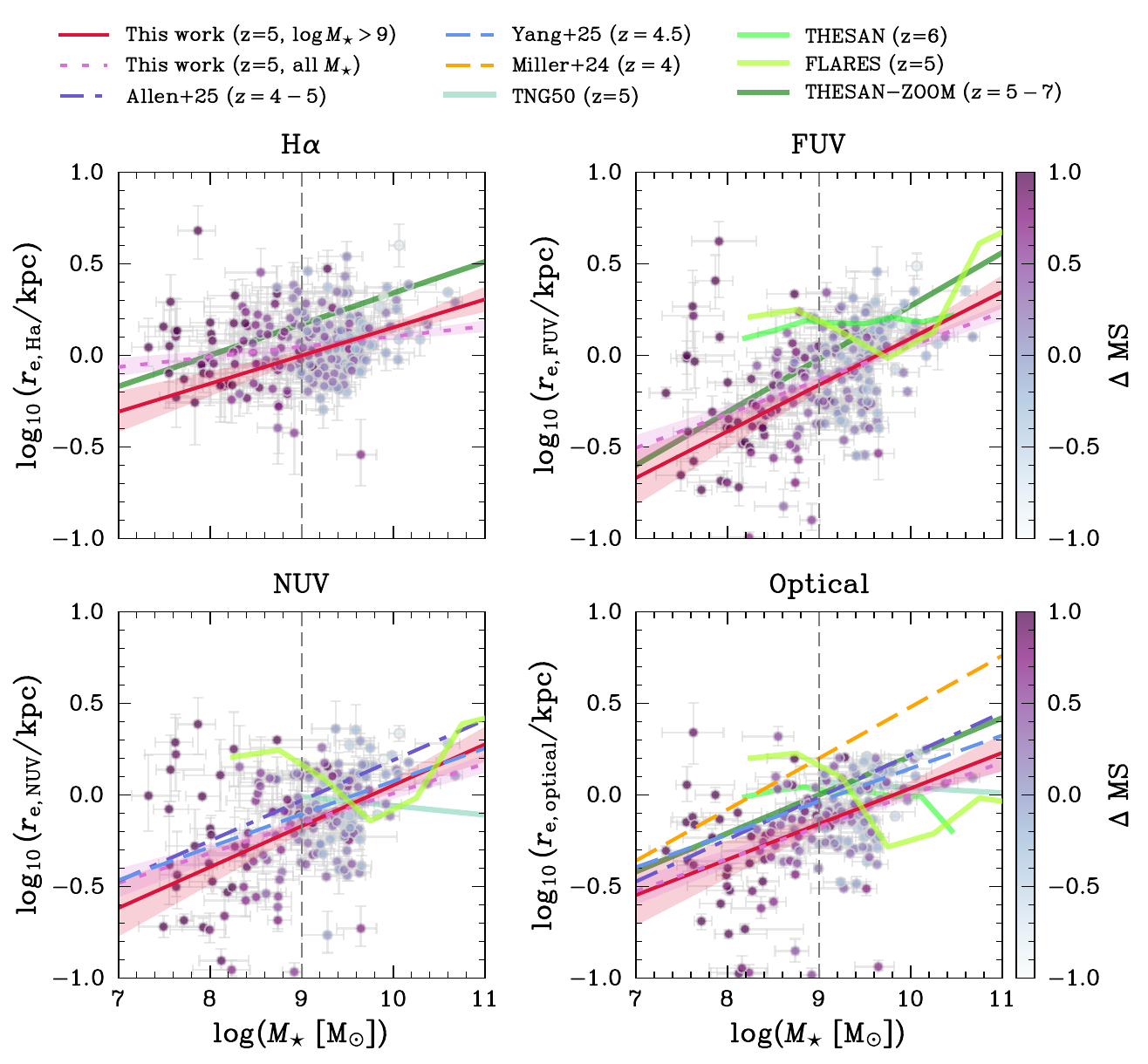}
    \caption{Size-mass relation at $z=4-6$ for sizes measured in $\Ha$ and rest-frame NUV, FUV, and optical for the galaxies in our sample (circles) colour-corded by their offset from the main sequence $\deltaMS$. We fit the SMR (Eq. \ref{eq:SMR}) for the galaxies above $\logMstar>9$, where our sample is representative, and plot the best-fit SMR at $z=5$ (red line). We also show our fit using the full sample (purple dashed line). The shaded regions for both fits show the uncertainties. We compare it to relations presented in the literature \citep{Miller:2024aa, Allen:2025aa,Yang:2025aa}, as well as prediction from cosmological simulations (\textsc{thesan}; \citet{Shen:2024aa}, \textsc{thesan-zoom}; \citet{McClymont:2025ab}, \textsc{tng50}; \citet{Costantin:2023aa}, and \textsc{flares}; \citet{Roper:2022aa}).}
    \label{fig:SMR-n}
\end{figure*}

\subsection{Size-mass relation (SMR)}

The SMR is typically fit by a single power law in the literature \citep[e.g.,][]{van-der-Wel:2014ab, Mowla:2019ab}. Following these works, we adopt a linear relation in logarithmic space, which accounts for both the redshift and the mass dependence of galaxy sizes:

\begin{equation}
\log \left(\frac{\Reff}{\rm kpc}\right) = \alpha \log \left(\frac{\Mstar}{10^{9.5} \rm \thinspace M_{\odot}}\right) + \beta + \gamma \log\left(\frac{1+z}{6}\right),
    \label{eq:SMR}
\end{equation}
where $\Reff$ is the half-light radius, $\Mstar$ is the stellar mass, and $\alpha$ is the slope of the mass dependence, $\beta$ is the normalisation (the size for a $\logMstar=9.5$ galaxy at $z=5$), and $\gamma$ parametrizes the redshift dependence. We favour this method over fitting a redshift-independent equation in different redshift bins \citep[e.g.][]{Allen:2025aa,Yang:2025aa} since our sample spans a relatively narrow redshift range, corresponding to a cosmic time interval of $\approx 0.6$ Gyr, and suffers from a decrease in galaxies at higher redshifts (see Fig. \ref{fig:mass-z}). Furthermore, fitting for both the mass and the redshift dependence simultaneously allows us to study the effect of each factor independently. This is difficult to do otherwise, since the average mass of galaxies naturally decreases with redshift. We also fit for the intrinsic scatter $\sigma_{\rm int}$ around the SMR, assuming that the sizes follow a log-normal distribution at fixed mass and redshift, $\mathcal{N}(1,\sigma_{\rm int}^2)$. 

In order to infer the best-fit parameters of the SMR, and their uncertainties, we use the Bayesian inference package \textsc{emcee} \citep{Foreman-Mackey:2013aa}. \textsc{emcee} samples the parameter space using an implementation of the affine invariant ensemble sampler for Markov chain Monte Carlo \citep{Goodman:2010aa}, resulting in posterior distributions for the model parameters. We adopt uniform priors for all parameters. The uncertainties in this work are defined with the $16^{\rm th}$ and $84^{\rm th}$ quantiles of the posterior distributions.

\renewcommand{\arraystretch}{2}
\begin{table*}
\centering
\begin{tabular}{c|c|c|c|c|c|c|c|c}
      & \multicolumn{2}{c|}{\Ha} & \multicolumn{2}{c|}{FUV} & \multicolumn{2}{c|}{NUV} & \multicolumn{2}{c}{Optical} \\ \hline
      & $\log M_{ \star }>9$ & all & $\log M_{ \star }>9$ & all & $\log M_{ \star }>9$ & all & $\log M_{ \star }>9$ & all \\ \hline
      $\alpha$   &   $0.15 \pm 0.04$   & $0.06 \pm 0.02$   & $0.25 \pm 0.06$ &   $0.19 \pm 0.03$ & $0.22 \pm 0.06$ &   $0.16 \pm 0.03$ & $0.19 \pm 0.07$ & $0.18 \pm 0.03$   \\ \hline
      $\beta$    &   $0.07 \pm 0.02$ & $0.07 \pm 0.02$ & $-0.03 \pm 0.02$ & $-0.04 \pm 0.02$&   $-0.06 \pm 0.02$& $-0.07 \pm 0.03$ &$-0.06 \pm 0.02$ & $-0.09 \pm 0.02$   \\ \hline
      $\gamma$   &   $-0.10 \pm 0.33$   & $-0.32 \pm 0.28$   &   $0.03 \pm 0.42$      & $-0.13 \pm 0.40$& $0.05 \pm 0.42$& $-0.31 \pm 0.40$ & $0.31 \pm 0.43$ & $-0.34 \pm 0.40$ \\ \hline
      $\sigma_{ \rm int }$ & $0.15 \pm 0.01$ & $0.16 \pm 0.01$ & $0.20 \pm 0.01$ & $0.24 \pm 0.01$ & $0.19 \pm 0.02$ & $0.24 \pm 0.01$ & $0.19 \pm 0.01$ & $0.23 \pm 0.01$ \\
\end{tabular}
\caption{Best-fit slope $\alpha$, normalisation $\beta$, and redshift dependence $\gamma$ for the SMR (Eq. \ref{eq:SMR})  probed by $\Ha$, FUV, FUV, and optical sizes. We quote here the fits done on galaxies with $\logMstar=9-10.5$, and on the full sample, $\logMstar=7.5-10.5$. The large uncertainties on $\gamma$ are likely due to the small redshift range probed, but the other parameters remain well constrained.}
\label{tab:size-fits}
\end{table*}

We present our results for the $\Ha$, FUV, NUV, and optical size-mass relation (Eq. \ref{eq:SMR}) at $z\approx 4-6$ in Fig. \ref{fig:SMR-n}, using our sample of 213 $\Ha$ emitters from the FRESCO and CONGRESS surveys. The $\Ha$ sizes are derived from the morpho-kinematic fitting of the grism data (Sec. \ref{sec:morph-kin-mod}), and the stellar continuum sizes are obtained from modelling with \textsc{pysersic} (Sec. \ref{sec:pysersic-mod}). We first fit the SMR only for galaxies with $\logMstar>9$, where our sample is most representative in terms of mass completeness ($\logMstar=9-10.5$). In fact, our selection bias comes from the integrated \Ha\ S/N limit, but since \Ha\ luminosity scales with stellar mass at fixed sSFR, this is approximately equivalent to a stellar mass cut. Above $\logMstar>9$, we find a flat trend between sSFR and stellar mass, suggesting completeness, whereas below we are biased to high sSFR systems. About half of our sample ($58\%$) is represented in the $\logMstar>9$ mass cut. 

In order to quantify the effects of this choice, we also include a fit done using the full sample. The best-fit parameters for the "fiducial" SMR and the one derived from the full sample are summarized in Tab. \ref{tab:size-fits}. We show the posterior distributions for the fiducial \Ha, FUV, NUV, and optical fits in Fig. \ref{fig:post-smr}. We plot both best-fit SMRs at $z=5$ on Fig. \ref{fig:SMR-n}, where we also compare to other SMRs at similar redshifts from the literature \citep{Miller:2024aa, Allen:2025aa,Yang:2025aa}. We also compare our observed SMR with predictions from simulations, namely \textsc{thesan} \citep{Kannan:2022aa,Garaldi:2022aa,Smith:2022aa}, presented in \citet{Shen:2024aa}, \textsc{thesan-zoom} \citep{Kannan:2025aa}, presented in \citet{McClymont:2025ab}, \textsc{tng50} \citep{Nelson:2019aa, Pillepich:2019aa}, presented in \citet{Costantin:2023aa}, and \textsc{flares} \citep{Lovell:2021aa,Vijayan:2021aa}, presented in \citet{Roper:2022aa}.

We find that the best-fit SMRs with and without the $\logMstar>9$ mass cut are consistent within the uncertainties for the FUV, NUV, and optical sizes, although the full sample slopes are all shallower by $\approx 0.05-0.17$ dex. This effect is stronger for the \Ha\ SMR which is almost flat when the full sample is included ($\alpha = 0.06 \pm 0.02$). The posterior distributions for the \Ha\ full sample fit are shown in Fig. \ref{fig:post-all}. This flattening could be partially caused by the large scatter of \sersic indices found at low masses, where we find preferentially higher values of $n$ (Fig. \ref{fig:n_ha}). Because of the degeneracy between $n$ and $\Reff$ (Eq. \ref{eq:Sersic_profile}), higher \sersic indices naturally result in larger sizes. Also, as highlighted in Fig. \ref{fig:mass-z}, most of the low mass galaxies $\logMstar<9$ in our sample lie above the main sequence, $\Delta \rm MS >0$, where 
\begin{equation}
    \Delta \rm MS = \log_{10}\left(SFR/SFR_{MS}(M_{\star})\right)
    \label{eq:deltams}
\end{equation}
and $\rm SFR_{MS}$ is the SFR, averaged over the last 10 Myr, for a galaxy of mass $M_{\star}$ on the main sequence. We adopt the SFMS from \citet{Simmonds:2025aa}. As we will discuss in Sec. \ref{sec:growth-and-kins}, this bias could cause the observed flattening of \Ha\ sizes if galaxies above the main sequence have boosted sizes. In the following sections, we focus on our fiducial fit, unless otherwise specified, but note that the main conclusions remain consistent with the full sample fit.

\subsection{Normalisation of the SMR}\label{sec:size-evol}

\begin{figure}
    \centering
    \includegraphics[width=0.90\linewidth]{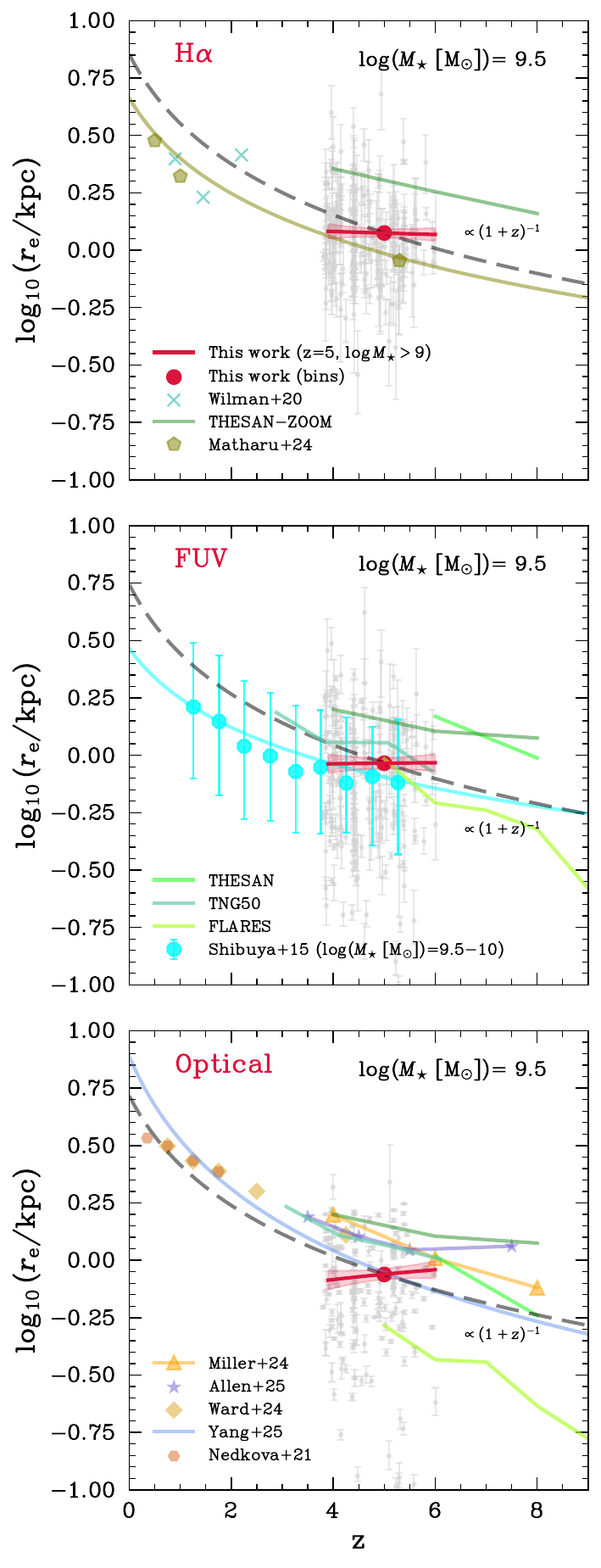}
    \caption{Average half-light ratio of $\Ha$ emission (top), FUV (middle), and optical continuum (bottom) for a $\lMstar=9.5$ galaxy as a function of redshift for the galaxies in this work (red circles). In all cases we find flat slopes for the redshift evolution (red lines; see Tab. \ref{tab:size-fits}). This is most likely due to the small redshift range probed, but could point to a flattening at high redshift. We compare with other works across cosmic time \citep{Shibuya:2015aa,Wilman:2020aa, Nedkova:2021aa,Matharu:2024aa,Miller:2024aa, Ward:2024aa,Allen:2025aa,Yang:2025aa}, who typically find negative slopes. This is also consistent with predictions from the \textsc{thesan}\citep{Shen:2024aa}, \textsc{thesan-zoom} \citep{McClymont:2025ab}, \textsc{tng50} \citep{Costantin:2023aa}, and \textsc{flares} \citep{Roper:2022aa} simulations.}
    \label{fig:SMR-Ha-z}
\end{figure}

We begin by focusing on the normalization of the SMR, parametrized by $\beta$, which we measure at $\logMstar = 9.5$ and $z=5$ (Eq. \ref{eq:SMR}). The $\Ha$ sizes, which probe the most recent star-formation ($t\approx 5-10 \rm \thinspace Myr$), are the largest, with $\rHa = 1.17 \pm 0.05$ kpc at $\logMstar = 9.5$. At this stellar mass, the stellar continuum sizes are all below 1 kpc, with $\rFUV = 0.93 \pm 0.05$ kpc and $\rNUV = \ropt = 0.87 \pm 0.05$ kpc. At all wavelengths, we see a large scatter of sizes at fixed stellar mass, spanning $\sigma \approx 0.5-1$ dex. The larger \Ha\ sizes are typically associated with the inside-out growth of galaxies, which we will discuss further in Sec. \ref{sec:mutliwave}, as they suggest that the recent star formation extends further out than the older stellar populations probed by the rest-frame UV and optical. Interestingly, we do not find a significant difference between UV and optical sizes, consistent with other studies from \textit{JWST} \citep{Treu:2023aa,Allen:2025aa,Miller:2024aa}. This could be due to a number of reasons. In terms of observational effects, outshining by young stellar populations could cause all of the continuum sizes to be probing stars of similar (young) ages. Also, dust attenuation can play a role, as we discuss in Sec. \ref{sec:dust}. Physically, the lack of difference in continuum sizes could imply that young stars are not preferentially forming in the outskirts of galaxies.

In Fig. \ref{fig:SMR-Ha-z}, we put our measured galaxy sizes in the context of other works from the literature. Compared to \citet{Matharu:2024aa}, who compute $\Ha$ sizes at similar redshifts based on stacks of $\Ha$ emitters in FRESCO, we find larger sizes by $\approx 0.1$ dex. Stacking suppresses features that are not perfectly aligned, so both small misalignments and structural variations could contribute to this discrepancy, particularly given that the maps are not corrected for kinematic distortions. Finally, although \citet{Matharu:2024aa} account for PSF effects in their morphological modelling, they do not take into account the grism LSF for the \Ha\ stacks, as we do in this work, which could also affect their size measurements.

We compare our rest-frame FUV size measurements to \citet{Shibuya:2015aa}, focusing on their sample of star-forming galaxies, probed in the UV out to $z\approx 6$. In the middle panel of Fig.~\ref{fig:SMR-Ha-z}, we show their median points for the $\logMstar = 9.5-10$ mass bin, whereas our results are plotted at fixed mass $\logMstar =9.5$. Nonetheless, our FUV average size $\rFUV = 0.93 \pm 0.05$ kpc is in good agreement with \citet{Shibuya:2015aa} at $z=5$. 

Turning to the optical sizes, plotted on the bottom panel of Fig.~\ref{fig:SMR-Ha-z}, we find excellent agreement with our measured median size at $z=5$ ($\ropt = 0.87 \pm 0.05$ kpc) and the sizes from \citet{Yang:2025aa} at this same redshift. Our median measurement lies below those from \citet{Ward:2024aa}, \citet{Allen:2025aa}, and \citet{Miller:2024aa}. The overall scatter in the reported median sizes from different observational studies points to a strong effect of the sample selection and modelling done in each work. Furthermore, a single \sersic modelling approach could be too crude, and is sensitive to the spatial resolution of the data, the sky background, and the contamination from neighbouring objects. The mass ranges probed in the different works are also important. We choose the reference mass $\logMstar = 9.5$ as it lies in a well constrained region of our sample, and fit galaxies with $\logMstar>9$. \cite{Allen:2025aa} probe galaxies down to $\logMstar \approx 8$, but as shown on Tab. \ref{tab:size-fits}, our median size is slightly smaller when we fit our full $\logMstar \approx 8-10$ mass range, so it does not help to explain the observed discrepancy. On the other hand, \citet{Ward:2024aa} only fits high-mass galaxies $\logMstar\geq 9.5$, so our reference mass lies at the cut-off of their sample. This is also the case for \citet{Miller:2024aa}, where they instead only fit low-mass galaxies $\logMstar\leq 9.5$. Finally, \citet{Yang:2025aa} use the same mass cut-off as this work and probes a similar range, which can help explain the consistency with our results at $z=5$.

Across wavelengths, we typically find smaller sizes than simulations. As shown in Fig. \ref{fig:SMR-Ha-z}, our sizes are $\approx 0.25$ dex smaller than those reported in \citet{McClymont:2025ab}. Notably, they only applied an observability cut based on continuum surface brightness, and so the discrepancy could be caused by observational selections which bias our sample towards more compact objects, missing more extended objects which at fixed SFR will have lower surface brightness. Importantly, the fainter emission at large radii seen in many simulations is typically missed due to the noise and PSF in observations, causing measured sizes to be smaller \citep{Punyasheel:2025aa}. This can, at least in part, explain why \textsc{thesan} \citep{Shen:2024aa} and \textsc{tng50} \citep{Costantin:2023aa} also lie above our median measurements in the FUV and optical. The \textsc{flares} median sizes for $\logMstar=9.5$ galaxies shown on Fig. \ref{fig:SMR-Ha-z} show good agreement with our measurements in the FUV, but lie well below in the optical. This is mainly due to the turnover in their SMR around $\logMstar=9.5$, as can be seen on Fig. \ref{fig:SMR-n}. Overall, their sizes are fairly consistent with our measurements in terms of normalization, but not mass dependence.

On the simulations side, there are various factors which affect the inferred sizes, starting with the chosen method to measure them (e.g. pixel- or aperture-based). Also, based on the strength of the implemented stellar feedback model, galaxies can be more or less bursty, which impacts the distribution of the stars and gas. For example, \textsc{thesan-zoom} is highly bursty. If the burstiness is too strong compared to observations, this could boost the observed sizes due to repeated strong feedback events. The dust content of the simulated galaxies also varies and can have an impact on the measured sizes, which we discuss in detail in Sec. \ref{sec:dust}.

\subsection{Slope of the SMR}\label{sec:slope-evol}
\begin{figure*}
    \centering
    \includegraphics[width=1\linewidth]{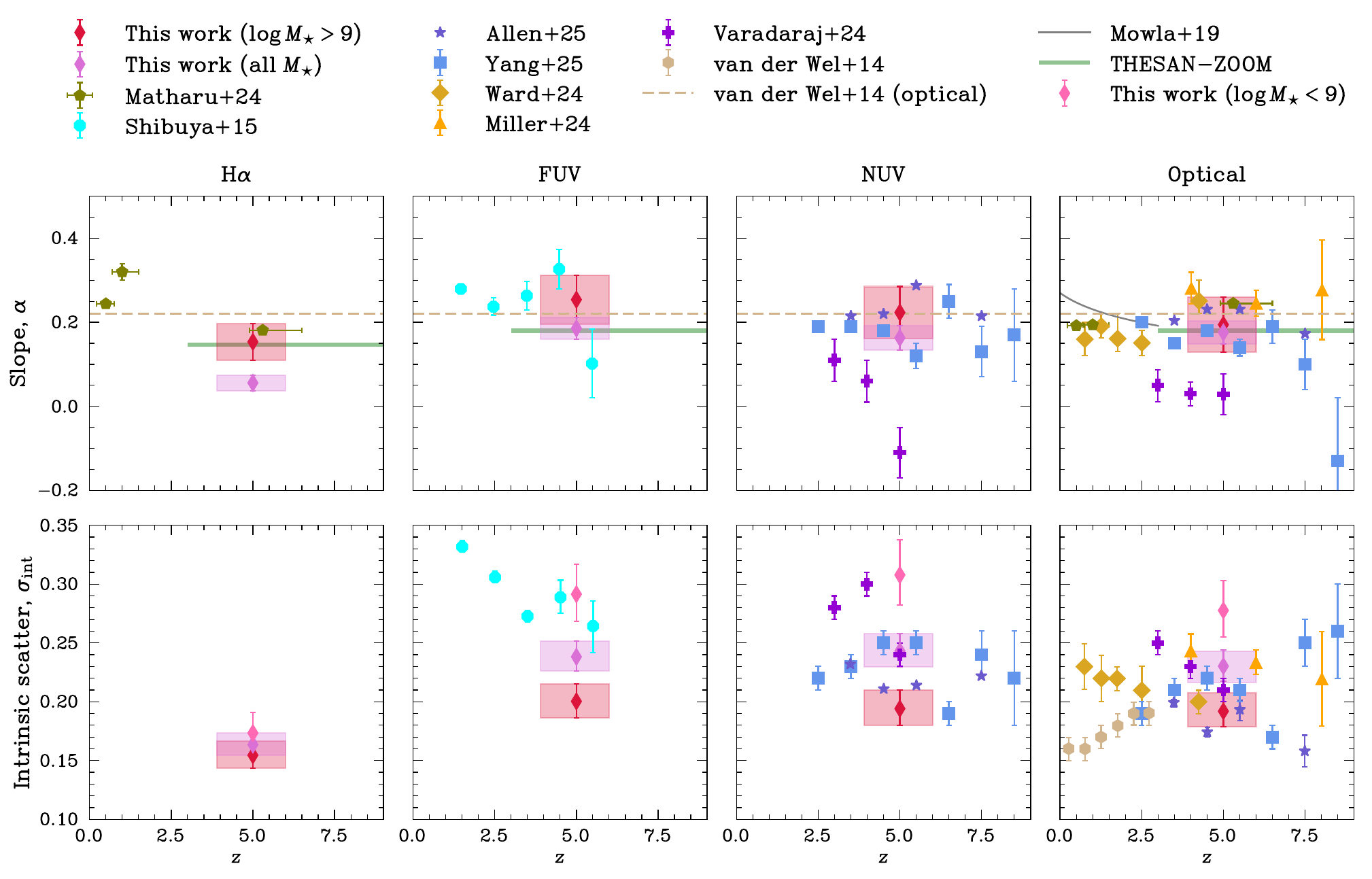}
    \caption{Evolution of the best-fit slope $\alpha$ of the mass dependence (top; Eq. \ref{eq:SMR}) and intrinsic scatter $\sigma_{\rm int}$ (bottom) for the \Ha, FUV, NUV, and optical sizes (left to right) for our sample with (red shaded regions) and without (purple shaded regions) applying a $\logMstar>9$ mass cut. We fit our full sample with a constant slope and scatter. We compare our measurements to other parametric fits from observations \citep{van-der-Wel:2014ab,Mowla:2019ab} and the \textsc{thesan-zoom} simulation \citep{McClymont:2025ab}, as well as results from fits in redshift bins across cosmic time \citep{Shibuya:2015aa,Matharu:2024aa,Miller:2024aa, Varadaraj:2024aa, Ward:2024aa, Allen:2015aa, Yang:2025aa}.}
    \label{fig:SMR-Ha-z-slope}
\end{figure*}

As shown in Fig. \ref{fig:SMR-n}, we find that galaxy sizes increase with mass at all probed rest-frame wavelengths (i.e. $\alpha >0$), although the slope and normalisation vary (Tab. \ref{tab:size-fits}). The \Ha\ sizes have a slightly weaker mass dependence ($\alpha = 0.15 \pm 0.04$) than the stellar continuum sizes ($\alpha \approx 0.19-0.25$), although the uncertainties are relatively large. \Ha\ sizes are sensitive to radiative transfer (RT) effects as they probe the extent of the ionization of the inter-stellar medium (ISM) by young massive stars. The shallower dependence on stellar mass could hence indicate that, at fixed mass, other properties such as SFR, can impact the size. 

We now place our measurements in the context of works at both lower and higher redshift in order to study the redshift evolution of the slope $\alpha$ of the SMR. On the top panel of Fig. \ref{fig:SMR-Ha-z-slope}, we plot our slopes at each wavelength, which are constant with redshift, for the $\logMstar=9-10.6$ and the full sample fits. We represent the uncertainty on the measurements, and the redshift span of our sample through the shaded regions. 

The shallower \Ha\ slope is consistent with \citet{Matharu:2024aa}, who find a steeper slope for the rest-frame optical continuum ($\alpha = 0.245 \pm 0.003$) than for the \Ha\ ($\alpha = 0.181 \pm 0.004$). In both cases, we find a slightly shallower slope (consistent within $1\sigma$), which could be attributed to stacking and LSF effects as discussed above. Interestingly, \citet{McClymont:2025ab} also find a steeper slope for the continuum compared to \Ha. These measurements, which are also consistent with ours within $1\sigma$ at all comparable wavelengths, suggest that there is an intrinsic, physically driven flattening of the \Ha\ mass dependence. We note that if \Ha\ is tracing non-circular motions such as radial inflows and outflows, this would also affect and perhaps flatten the SMR.

The slopes for the continuum suffer from larger uncertainties, but are consistent with $\alpha \approx 0.2$ within $1\sigma$. We compare our results in the FUV with \citet{Shibuya:2015aa}. We note that in their work, the growth of sizes is based on UV luminosity $L_{\rm UV}$ and not mass, so this is the slope reported on Fig. \ref{fig:SMR-Ha-z-slope}. This should offer a fair comparison given the linear relation between $\Mstar$ and $L_{\rm UV}$ at fixed redshift \citep[e.g.][]{Daddi:2007aa}. Our measured slope of $\alpha = 0.25 \pm 0.06$ at $z=5$ lies between the measurements from \cite{Shibuya:2015aa} above and below $z=5$, although consistent within the uncertainties. 

In the rest-frame NUV and optical sizes, we find good agreement with other surveys at similar redshifts \citep{Miller:2024aa, Ward:2024aa, Allen:2025aa, Yang:2025aa}. \citet{Varadaraj:2024aa} find much flatter slopes ($\alpha \approx 0.04$) at $z\approx3-5$, which could be due to their ground-based sample selection, which is biased towards luminous systems. In the context of lower redshift measurements \citep{Matharu:2024aa,Ward:2024aa}, we find evidence for a constant slope of the SMR. This is consistent with predictions from \citet{van-der-Wel:2014ab} who find a constant slope $\alpha=0.22$ at $0<z<3$. We do not find evidence for a change in slope for the FUV and NUV sizes in our sample, which are in agreement with the optical slope within $1\sigma$. This is consistent with the results reported in \citet{Allen:2025aa,Yang:2025aa}. The lack of wavelength dependence of the slope implies that dust attenuation is weak at these redshifts, or at least has no strong mass dependence. When comparing with the \textsc{tng50} \citep{Costantin:2023aa}, \textsc{thesan} \citep{Shen:2024aa}, and \textsc{flares} simulations \citep{Roper:2022aa}, our observed positive SMR is in tension with the negative or flat slopes reported in these works, as can be seen on Fig. \ref{fig:SMR-n}. We plot the "observed" SMRs from the simulations, meaning that they are sensitive to dust attenuation effects in the same way as observations, and should offer a more direct comparison. We discuss this further in Sec. \ref{sec:dust}.

\subsection{Redshift dependence of the SMR}\label{sec:z-evol}

We now focus on the redshift dependence of the SMR, parametrized by $\gamma$ in Eq. \ref{eq:SMR}. We focus on the rest-frame optical and FUV only, from the continuum sizes, since they offers the most comparison samples from the literature. We plot our best-fit redshift dependence on Fig. \ref{fig:SMR-Ha-z}, as a solid red line, for galaxies at $\logMstar = 9.5$. It is important to note that we are not plotting or discussing the evolution of one galaxy, but rather the evolution of the population of $\logMstar = 9.5$ galaxies across cosmic time. Each galaxy will presumably only have a stellar mass of $\logMstar = 9.5$ once in its mass assembly history, and we are instead interested in studying whether the average size of a $\logMstar = 9.5$ galaxy changes with time. Galaxies with $\logMstar = 9.5$ at $z=5$ will have evolved into massive ($\logMstar\approx 10-11$) galaxies at lower redshift, and might even have ceased their star formation. 

Starting with \Ha, we find a negative but weak redshift dependence $\gamma = -0.10\pm0.35$, which suffers from large uncertainties due to the small redshift range probed in this work. This implies that the average size of a $\logMstar = 9.5$ galaxy remains at $\rHa \approx 1.2$ kpc at $z\approx 4-6$. Comparing to studies at $z<3$ \citep{Nelson:2016wo,Wilman:2020aa, Matharu:2024aa}, we find that the average size decreases by $\approx 0.15$ dex from $z=2$ to $z=5$. In order for \Ha\ sizes at $\logMstar = 9.5$ to evolve from $\rHa = 1.2 \pm 0.05$ at $z=5$ to $\rHa = 3.16$ kpc at $z=0.5$ \citep{Nelson:2016wo,Matharu:2024aa}, they would have to follow a redshift evolution of $\rHa\propto (1+z)^{-0.66}$, stronger than the one measured in our sample alone, $r_{\rm e, \Ha}\propto (1+z)^{-0.12}$. This points towards a different redshift dependence at high and low redshift, which we discuss more in Sec. \ref{sec:disc-SMR}. Nonetheless, the trend shown on Fig. \ref{fig:SMR-Ha-z} implies that a $\logMstar=9.5$ galaxy at $z=5$ will be more compact than a $\logMstar=9.5$ galaxy at $z=2$.

For the stellar continuum sizes, we find a mild positive slope which is consistent with no evolution. Specifically, we infer $\gamma = 0.31\pm0.43$ for the optical sizes, $\gamma = 0.03\pm0.42$ for the FUV, and $\gamma = 0.05\pm0.42$ for NUV. This supports the same conclusion as discussed for the \Ha\ sizes, namely a flattening of the redshift dependence of the SMR at high redshift. Other observational studies predict a range of different slopes. When comparing \citet{Yang:2025aa} and \citet{Ward:2024aa}, neither slope is able to reconcile our measurement at $z\approx 5$ with the sizes from \citet{Nedkova:2021aa} at $z<2.5$. \citet{Miller:2024aa} also discuss the need for a redshift dependent slope, although they argue the slope increases at high-redshift. They also test a parametrization as a power law in the Hubble constant, $r_e\propto H(z)^{\gamma}$, and are still not able to reconcile measurements at low and high redshift with a constant slope $\gamma$. 

Differences in the derived redshift evolutions could be attributed to variations in the mass dependence $\alpha$, as well as samples probing different mass and redshift ranges and hence being more or less sensitive to the redshift evolution at $\logMstar = 9.5$. However, it is also important to note that we fit the SMR within our full sample, accounting for the mass and redshift evolution independently. In other works, such as \citet{Ward:2024aa} and \citet{Allen:2025aa}, the redshift evolution is inferred from median sizes (at fixed mass) in different redshift bins. This subtle difference could influence the median sizes inferred, as this latter method does not take into account redshift evolution within a specific bin, and could hence be influenced by bin size, for example.

\citet{McClymont:2025ab} argue for the need of a double power-law fit for the sizes in the \textsc{thesan-zoom} simulations, finding a shallow slope at the redshifts probed here. However, exploring a more complex redshift evolution is beyond the scope of this work, as our sample does not span a large enough parameter space. On Fig. \ref{fig:SMR-Ha-z}, we show in grey the evolution $r\propto (1+z)^{-1}$ as expected for the virial radius $R_{\rm vir}$. We fix the normalization to the median sizes at $z=5$ measured in this work for \Ha\, FUV, and optical. Qualitatively, for $\logMstar =9.5$ galaxies, the size evolution does not appear fully consistent with the expected growth of haloes. This suggests that the relation between the sizes of galaxies and their host haloes varies with redshift, as is expected for the relation between stellar and halo masses, traced by the stellar-to-halo-mass (SMHM) relation \citep{Behroozi:2019aa}.

\subsection{Evolution of the intrinsic scatter}\label{sec:scatter-evol}

Finally, we discuss our results regarding the intrinsic scatter, $\sigma_{\rm int}$, of galaxies around the SMR. This scatter sheds light on whether other physical processes influence the sizes of galaxies at fixed stellar mass. We plot our best-fit measurements on the bottom panel of Fig. \ref{fig:SMR-Ha-z-slope}. 

The scatter for all of the continuum sizes is consistent within the uncertainties, $\sigma_{\rm int} \approx 0.19-0.20$, and is larger than its counterpart for the \Ha\ sizes, $\sigma_{\rm int} = 0.15\pm 0.01$. Across all wavelengths, the scatter increases when the full sample is fit. This implies that the scatter increases at lower masses, as can be directly seen on Fig. \ref{fig:SMR-n}. Although this could be due to a number of reasons, it is likely caused by smaller sizes being more difficult to constrain due to the size and complex shape of the PSF. These measurements therefore suffer larger uncertainties. Also, the \sersic index is more difficult to constrain for smaller systems, and tends to be higher than for larger more massive galaxies (see Fig. \ref{fig:n_ha}). Because the \sersic index is degenerate with the effective radius, this affects the best-fit sizes and introduces more variation.

In the bottom panel of Fig. \ref{fig:SMR-Ha-z-slope}, we also plot measurements $\sigma_{\rm int}$ from other studies at different redshifts. In the NUV and optical, the values from the literature vary between $\sigma_{\rm int} = 0.15-0.25$, but the scatter seems to remain at around $\sigma_{\rm int} = 0.2$ independently of redshift, consistent with the scatter measured from our fiducial fits. This scatter is however difficult to constrain as it strongly depends on understanding the measurement uncertainties, which can be derived in different ways in the literature. In our work, they are derived in a fully Bayesian framework and marginalized over all of the free parameters. Nonetheless, the relative agreement shown in Fig. \ref{fig:SMR-Ha-z-slope} suggests that the measured scatter has a distinct physical nature, which we discuss in Sec. \ref{sec:disc-sizes}. In the FUV, \citet{Shibuya:2015aa} find a significantly larger scatter from $z=1-6$. We also note that their scatter is defined as the standard deviation of the log-normal distribution of sizes at fixed $L_{\rm UV}$, so we convert it to a decimal (base 10) logarithm to include it in our comparison. They do not fit for the intrinsic scatter but instead account for the measurement errors in quadrature, which could in part explain the discrepancy with our measurements. Importantly, as mentioned, they measure the scatter in the $r-L_{\rm UV}$ relation, and not the $r-\Mstar$ relation, which could introduce systematics. 

\section{Multi-wavelength sizes and inside-out growth} \label{sec:mutliwave}
In this section, we utilize our multi-wavelength size measurements to study how their ratios depend on different physical quantities. Specifically, in Sec. \ref{sec:insideout}, we quantify the dependence of the ratio of \Ha-to-UV sizes on the position of galaxies above or below the main sequence, and discuss our results in the context of inside-out growth. We also investigate the compaction of galaxies in Sec. \ref{sec:compaction} by disentangling the mass and $\deltaMS$ dependence of their FUV sizes. In Sec. \ref{sec:dust}, we instead focus on the ratio of FUV-to-NUV sizes to explore the effects of dust attenuation in our sample. 

\subsection{Inside-out growth, or the lack thereof} \label{sec:insideout}

\begin{figure*}
    \centering
    \includegraphics[width=1\linewidth]{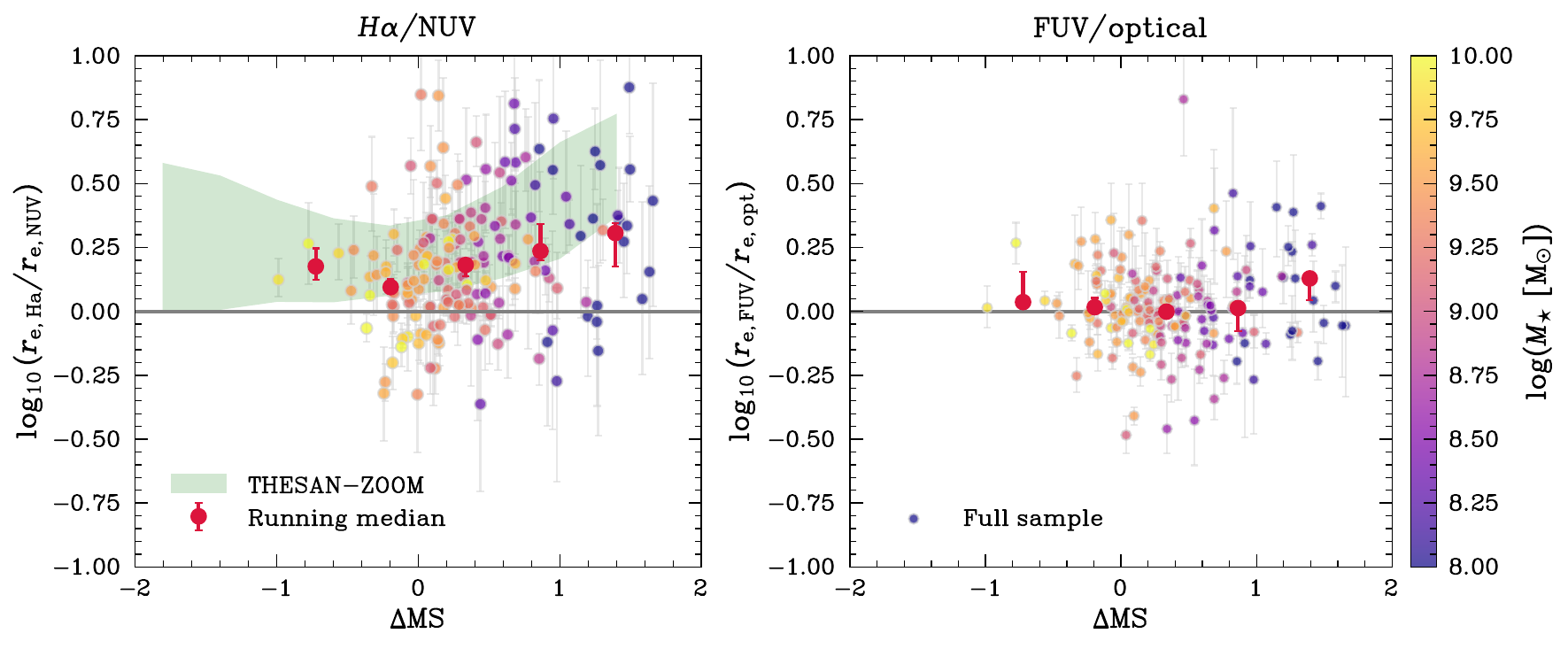}
    \caption{(left) Dependence of the ratio of $\Ha$ to NUV sizes with the offset from the main sequence $\deltaMS$, on a 10 Myr timescale, for our full sample (color-coded by stellar mass) and running medians (red circles). Most of our sample lies in the $\log(\HatoUV) > 0$ regime, meaning the $\Ha$ is more extended than the UV emission. The \textsc{thesan-zoom} simulation \citep[green shaded region,][]{McClymont:2025ab} predict an increase of $\HatoUV$ above the main sequence, which we find evidence for in our sample (red circles). The width of the shaded region represents the scatter in the underlying distribution. (right) Evolution of the ratio of FUV to optical sizes with the offset from the main sequence $\deltaMS$ for our full sample (color-coded by stellar mass) and running medians (red circles). As the FUV and optical continuum trace young and older stellar populations, respectively, their size ratio probes inside-out versus outside-in growth. We find no trend within our sample, which suggests that the extended \Ha\ sizes are not driven by inside-out growth but by radiative transfer effects.}
    \label{fig:HaUV-deltaMS}
\end{figure*}

In numerous studies at lower redshifts, the ratio of \Ha\ to rest-frame UV or optical sizes has been used to motivate the inside-out growth of galaxies \citep[e.g.,][]{Nelson:2016wo,Wilman:2020aa,Matharu:2024aa}. \Ha\ emission is driven by the ionization of gas by young massive stars, through the emission of LyC photons, and hence traces star formation on short timescales ($\approx 10$ Myr), reflective of these stars' lifetime. The rest-frame UV traces star formation on longer timescales ($\sim 10-100$ Myr), and finally the rest-frame optical is a good tracer of the galaxy's integrated star-formation history. Studying the ratio of galaxy sizes at these different wavelengths provides insight on how star formation has evolved spatially within the galaxy. At higher redshifts, however, galaxies are younger, more gas-rich, and often dominated by recent bursts of star formation. In such systems, strong outshining from young stellar populations and rapid structural evolution may alter the relative sizes measured at different wavelengths, potentially leading to different interpretations than at lower redshifts ($z<3$).

In the left panel of Fig. \ref{fig:HaUV-deltaMS}, we plot the ratio of \Ha\ to NUV sizes as a function of the galaxy's offset from the main sequence, $\Delta\rm MS$. Most of our sample ($77\%$) has larger \Ha\ sizes compared to the NUV, $\log (\HatoUV) > 0$. In terms of star-formation timescales, this implies that younger stars are forming in the outskirts, beyond where the earlier generation had formed. However, following this logic, we should see the same trend when looking at the ratio of FUV to optical sizes. We plot $\FUVtoopt$ as a function of $\deltaMS$ on the right panel of Fig. \ref{fig:HaUV-deltaMS}. As highlighted by the running medians, we find that our sample is scattered around $\log (\FUVtoopt) = 0$, providing no evidence for steady inside-out growth. Based on Fig. \ref{fig:SMR-Ha-z}, the shallow increase of average sizes at fixed mass, at least at high redshift, would also support this statement. This implies that galaxies could be growing inside-out on average, but both the optical and UV are dominated by the last burst.

This result suggests that the larger \Ha\ sizes, compared to the UV, are probably explained by another phenomenon. The left panel of Fig. \ref{fig:HaUV-deltaMS} shows an increase of \HatoNUV above the main sequence, $\deltaMS>0$. In fact, we find a mild but significant correlation between these two quantities ($\rho = 0.27$, $p<0.001$), and see a clear increase in the running medians from $\log(\HatoNUV) = 0.1$ to $\log(\HatoNUV) = 0.3$. We also find evidence for a slight increase below the main sequence, although our sample size drops significantly in that region.  When comparing to the \textsc{thesan-zoom} simulations \citep{McClymont:2025ab}, we find good agreement with this observed trend. As discussed in \citet{McClymont:2025ab}, the boost in \Ha\ sizes above the main sequence can instead be explained by large amounts of LyC photons produced during bursts of star formation, which increases the radii of their Strömgren sphere \citep{Stromgren:1939aa}. The Strömgren sphere describes the ionizing front of the LyC photons, which becomes the radial extent of the induced \Ha\ emission. Hence, the \Ha\ sizes do not directly trace their source of ionization, i.e., the young massive stars, but rather reflect the intensity of the ongoing star formation. During a burst, where $\deltaMS$ is high, the \Ha\ sizes will then naturally be at their highest. Given the lack of boost of NUV sizes compared to the optical, this phenomenon is a more likely explanation for the larger \Ha\ sizes than inside-out growth. This is supported by observations finding that star formation is bursty at high redshift \citep[e.g.][]{Looser:2023aa,Tacchella:2023aa,Baker:2025aa,Simmonds:2025aa}, which promotes galaxy growth through central starbursts \citep{Tacchella:2016aa,El-Badry:2016aa, McClymont:2025ab} rather than a star-forming disk. We find evidence for this in our work, with the observed increase of $\HatoNUV$ with $\deltaMS$ is in part driven by the decrease of UV sizes during bursts (Fig. \ref{fig:rUV-deltaMS}). The results are also consistent with medium-band imaging observations, which show that the relative size of emission line regions compared to the continuum scales with the ionizing photon production rate \citep{Zhu:2025aa}.

To qualitatively explore the validity of our Strömgren sphere interpretation, we estimate the Strömgren radii $R_S$ for the galaxies in our sample using a simple approximation. The Strömgren radius is empirically defined as $R_S = (\frac{3S_{\star}}{4\pi n^2 \beta})^{1/3}$, where $S_{\star}$ is the ionising flux, $n$ is the hydrogen number density, and $\beta$ is the total recombination rate. Assuming a Case B recombination rate for an inter-stellar medium (ISM) temperature of $10^4$ K, and an average density of $10$ cm$^{-3}$, we can estimate $R_S$ using the integrated \Ha\ flux, scaled by the redshift to obtain the intrinsic luminosity. We find a median $R_S\approx 0.8$ kpc, which is comparable to our measured \Ha\ radii, meaning our interpretation is plausible.

Our finding of large \Ha\ to NUV size ratios at $z\sim 5$ is consistent with measurements from Villanueva et al (in prep.) at $z=4.4-4.9$, where they model and subtract the rest-optical continuum from the F356W photometry, by using non-contaminated neighbouring medium band filters, to obtain \Ha\ maps. They attribute the large \Ha\ sizes to a combination of vigorous star formation and leaking ionising photons. Interestingly, the recent study from the \textit{JWST} Emission Line Survey \citep[JELS;][]{Stephenson:2025aa} does not find that the nebular (\Ha) sizes are larger than the stellar continuum (UV or optical) sizes on average. In \citet{Stephenson:2025aa}, the \Ha\ emission is traced by narrow band imaging for 23 star-forming galaxies at $z=6.1$, and is actually found to be typically smaller than the stellar continuum traced by the rest-frame $R$-band (F444W), with a median $r_{\rm e, F444W}/r_{\rm e, NB} = 1.20 \pm 0.09$. The small sample size and large scatter in the measurements, which could be in part caused by the low stellar masses probed ($\logMstar\approx8-9.5$), may be responsible for the discrepancy with our results. Furthermore, their selection is based on narrowband H-alpha emission which could cause a bias towards galaxies with centrally concentrated profiles. Discrepancies could also be caused by the different modelling approaches, where \citet{Stephenson:2025aa} fit the narrow-band imaging, containing the \Ha\ emission, with \textsc{galfit} and we forward-model the \Ha\ emission line in the grism data.

\subsection{Evidence for compaction above the MS} \label{sec:compaction}

\begin{figure}
    \centering
    \includegraphics[width=1\linewidth]{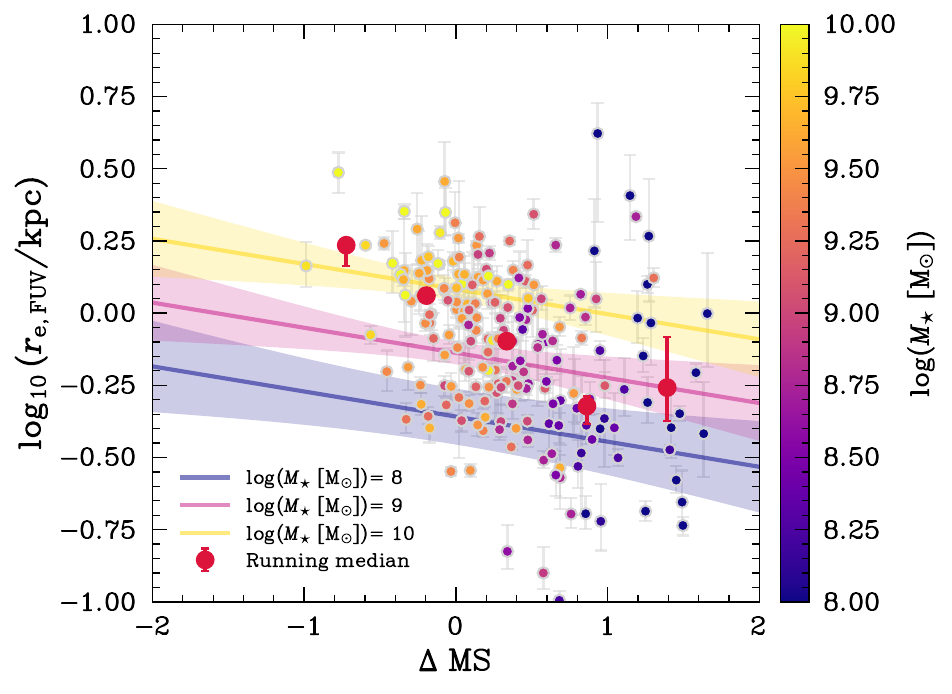}
    \caption{Dependence of the FUV sizes $\rFUV$ on the offset from the main sequence $\deltaMS$. The running medians (red points) show a clear decrease of sizes with $\deltaMS$. Due to the multi-dependence of sizes on both $\Mstar$ and $\deltaMS$ in our sample, we plot our best-fit $r-\deltaMS$ relation at fixed redshift ($z=5$) and mass ($\logMstar = 8-10$) in solid lines. These relations show the residual dependence of sizes on $\deltaMS$ once the dependence of stellar mass is accounted for.}
    \label{fig:rUV-deltaMS}
\end{figure}

A key factor to consider when studying the dependence of multi-wavelength size ratios with $\deltaMS$ (Fig. \ref{fig:HaUV-deltaMS}) is the behaviour of the individual sizes themselves. We investigate the rest-frame FUV sizes, as they are expected to probe star formation on the shortest timescales. As shown on Fig. \ref{fig:rUV-deltaMS}, we find a clear decrease of $\rFUV$ with $\deltaMS$, suggesting that galaxies go through a compaction as they move above the main sequence. However, selection effects lead to a form of degeneracy between stellar mass and offset from the main sequence in our sample (Fig. \ref{fig:mass-z}), which means the distinct effects of these two quantities become intertwined. We attempt to break this degeneracy using our Bayesian fitting approach with \textsc{emcee}.

As described in Sec. \ref{sec:res-SMR}, we fit our SMR by also accounting for a redshift dependence (Eq. \ref{eq:SMR}). To quantify the dependence of sizes on both stellar mass and $\deltaMS$, we compute a new fit of the SMR where we further include a power-law dependence on \deltaMS by adding a $\delta\cdot \deltaMS$ term in Eq. \ref{eq:SMR}, since $\deltaMS$ is already in log-space (Eq. \ref{eq:deltams}). We show the posterior distributions for $\alpha$, $\beta$, $\gamma$ (Eq. \ref{eq:SMR}) and the new parameter $\delta$ on Fig. \ref{fig:post-deltams}. Compared to the fit without the $\deltaMS$ dependence (Tab. \ref{tab:size-fits}), we find that the mass dependence is slightly weaker in this new fit, from $\alpha = 0.25 \pm 0.06$ to $\alpha = 0.22 \pm 0.06$, although consistent within the uncertainties. More importantly, we find a weak positive, non-zero dependence of $\rFUV$ on $\deltaMS$, $\delta = 0.08 \pm 0.06$. This implies that within our data, the FUV sizes decrease above the main sequence independently of their stellar mass or redshift. 

We plot the best fit $r-\deltaMS$ relation at fixed redshift ($z=5$) and for a range of stellar masses ($\logMstar = 8,9,10$) on Fig. \ref{fig:rUV-deltaMS}. As expected, we find the intrinsic $\deltaMS$ dependence to be weaker than the trend traced by the medians, meaning some of the observed decrease is driven by stellar mass. Larger, more complete samples are needed to better quantify the dependence of sizes on $\Mstar$ and $\deltaMS$. Our running medians provide a form of upper limit on the strength of the  $r-\deltaMS$ relation, since they implicitly assume that \deltaMS is the only factor. When we account for mass in our sample, we find that the dependence on \deltaMS is weaker, but this could be underestimated due to a biased sample at low masses.

The compaction of galaxies as they move up the main sequence has been studied in cosmological simulations \citep{Zolotov:2015aa,Tacchella:2016aa, Lapiner:2023aa,McClymont:2025ab}. This state is characterised by compact, gas-rich cores of star formation which occur when the galaxy reaches its peak star formation. This strong, centrally concentrated star formation then leads to temporary inside-out quenching where feedback drives the gas from the core \citep{Tacchella:2016aa,McClymont:2025aa,McClymont:2025ab}. These oscillations between compaction and quenching predict that $\deltaMS$ should influence the observed sizes of galaxies independent of their mass, with galaxies above the main sequence being seen preferentially in compaction phases. In this work, we find evidence of this trend, despite significant scatter and uncertainties. Studying where quiescent galaxies, which lie well below the main sequence, lie on these trends would be highly informative but is beyond the scope of this work since we are limited by the need for \Ha\ detections. In this work, we only probe systems that are in the process of being temporarily quenched, meaning they have falling SFHs. 

\subsection{Central dust attenuation}\label{sec:dust}

Even though most observations find positive size-mass relations across redshifts (Sec. \ref{sec:slope-evol}), cosmological hydrodynamical simulations have struggled to reproduce this result as they find flat or negative slopes at high redshifts \citep{Marshall:2022aa,Roper:2022aa,Costantin:2023aa, Shen:2024aa}, as shown in Fig. \ref{fig:SMR-n}. This discrepancy is often attributed to the effects of dust attenuation, which increases with stellar mass and is stronger in the central regions of the galaxy \citep{Wu:2020aa,Marshall:2022aa,Popping:2022aa,Roper:2022aa,Costantin:2023aa}. When they take into account dust, they are able to increase their slope because the dust boosts the sizes of massive galaxies by obscuring the central regions. \citet{Punyasheel:2025aa} show that when accounting for observation effects, such as noise and PSF, using the FLARES simulations, the slope then even becomes positive. 

Recently, \citet{McClymont:2025ab} found an intrinsically positive size-mass relation in the \textsc{thesan-zoom} simulations, and argued that this was recovered due to the bursty nature of star formation in those simulations allowing for regular compaction and expansion as galaxies build up mass. This aligns with the findings of \citet{Roper:2023aa}, who claim that the negative size-mass relation at high redshift in large-volume cosmological simulations with an effective equation-of-state ISM model is due to runaway gas collapse down to the gravitational softening length in the most massive, highly star forming galaxies at high redshift in these models. This raises the interesting prospect that we can constrain models of galaxy formation by observationally measuring the impact of dust on high-redshift galaxy sizes.

\begin{figure}
    \centering
    \includegraphics[width=1\linewidth]{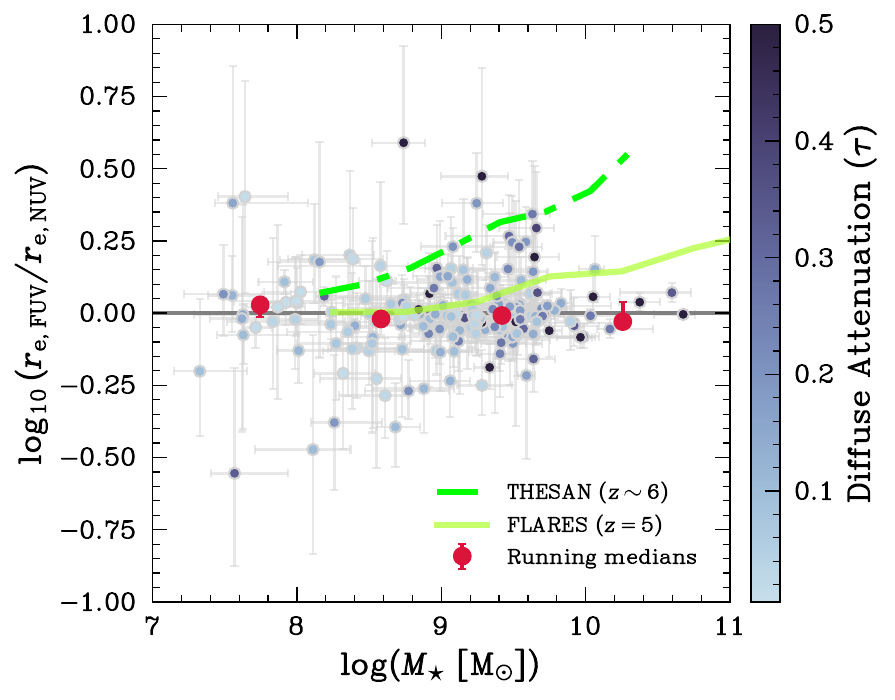}
    \caption{Dependence of the ratio of FUV-to-NUV sizes as a function of stellar mass. The points are colour-coded by the optical depth of the diffuse dust attenuation in the V-band $\tau_V$ derived from the \textsc{Prospector} fits. As shown by the low values $\tau_V\approx 0.2$ and the running medians centred on $\log_{10}(r_{\rm e, FUV}/r_{\rm e, opt}) = 0$, we do not find evidence for a mass-dependent trend of central dust attenuation in our sample at $z\approx4-6$. We qualitatively illustrate the expected effects of dust by comparing to the \textsc{thesan} simulations \citep[dashed line;][]{Shen:2024aa}, with the ratio of observed UV to intrinsic UV (no dust) sizes. We also plot the FUV-to-NUV sizes from the \textsc{flares} simulations \citep[solid line;][]{Roper:2022aa}, where the effects of dust are visible at the high mass end.}
    \label{fig:dust-mass}
\end{figure}

In order to investigate whether our recovered slopes of $\alpha=0.15-0.25$ could be due, at least in part, to dust attenuation, we show the ratio of FUV to NUV sizes in Fig. \ref{fig:dust-mass} as a function of stellar mass. We colour-code our galaxies by the measured dust attenuation, extracted from our SED modelling results, which is represented by the optical depth $\tau_V$. These two rest-frame wavelength ranges probe similar physical processes at high redshift, but the bluer wavelengths (FUV) are more affected by dust attenuation. If our galaxies were affected by central dust obscuration, we would expect to see a positive value of \FUVtoopt, as the FUV radii would appear larger. This effect should be more extreme for higher mass galaxies, which typically have more dust. However, as highlighted by the running medians on Fig. \ref{fig:dust-mass}, our sample is scattered around $\log(\FUVtoNUV) = 0$, showing no systematic enhancement of FUV sizes. The scatter around $\log(\FUVtoNUV) = 0$ is consistent with measurement errors and the intrinsic scatter measured around the SMR (Fig. \ref{fig:SMR-Ha-z-slope}, bottom panel). The measured attenuations from SED modelling with \textsc{prospector} show that most of our galaxies have relatively low dust attenuation ($\tau_V\leq 0.2$), consistent with the lack of observed mass trend. Some of our systems appear to have more dust, with $\tau_V\approx 1$. However, as shown in \citet{Maheson:2025aa}, the dust attenuation in the V-band is strongly correlated with stellar mass, which would introduce a bias in our SMR as suggested by simulations. Crucially, we find no trend of $\FUVtoNUV$ with stellar mass, as shown by the flat running medians, implying our positive SMR slope is not caused by dust attenuation effects, despite some galaxies potentially having non-negligible dust contents.

We compare our measurements to the FUV-to-NUV ratio measured in the \textsc{flares} simulations \citep{Roper:2022aa}. As can be seen on Fig. \ref{fig:dust-mass}, there is a trend of increasing \FUVtoNUV\ with stellar mass, although our sample is scarce at the high masses where this becomes significant in \textsc{flares}. Nonetheless, this boosting at high masses is consistent with the galaxies having significant central dust obscuration. It highlights that the intrinsic SMRs would be strongly affected by the increasing dust attenuation for higher mass galaxies. We qualitatively also compare to the \textsc{thesan} simulations \citep{Shen:2024aa} by plotting the ratio of observed UV to intrinsic UV (no dust) sizes. This is not an apples-to-apples comparison, but it illustrates the effects of dust attenuation well. The lack of such a trend in our data suggests that our measured SMR is intrinsically positive and not affected by dust. These comparable NUV and FUV sizes are also found in Villanueva et al. (in prep.) at similar redshifts ($z=4.4-4.9$).

\section{Galaxy growth through kinematics} \label{sec:growth-and-kins}

In this section, we investigate how the link between morphology and kinematics can help paint a picture of galaxy growth in the early Universe. In Sec. \ref{sec:kins}, we discuss the evolution of the ratio of \Ha-to-NUV sizes with the ionized gas rotational support $\rotsupp$, and in Sec. \ref{sec:morph-disks}, we instead link $\rotsupp$ to its commonly used morphological "counterpart", the observed axis ratio $b/a$.

\subsection{Kinematics along the MS}\label{sec:kins}

\begin{figure}
    \centering
    \includegraphics[width=1\linewidth]{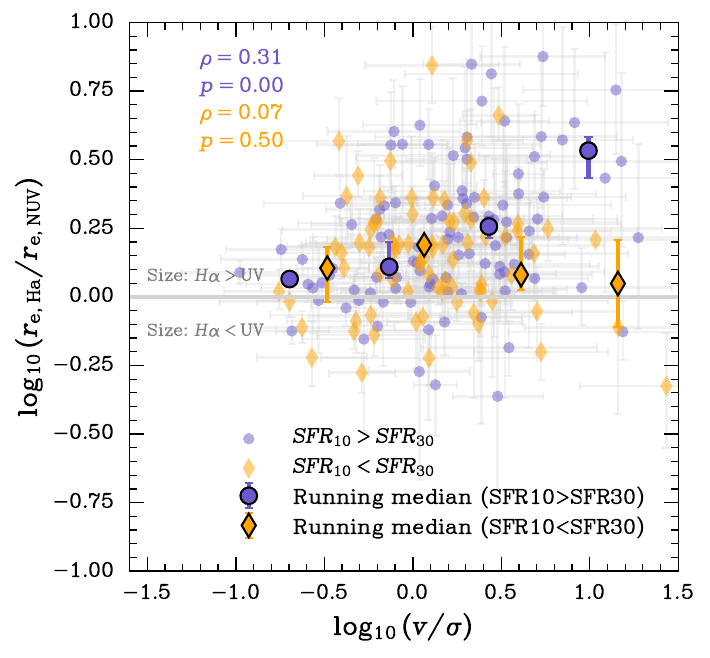}
    \caption{Dependence of $\HatoUV$ with the rotational support $\vsigma$ for galaxies with $\rm SFR_{10}>SFR_{30}$ (i.e., galaxies moving up the main sequence, purple) and $\rm SFR_{10}<SFR_{30}$ (i.e., galaxies moving down the main sequence, orange). Whereas we find no trend for the galaxies moving down, we find a mild correlation for galaxies moving up.  This suggests that as galaxies move up the main sequence, they are accreting gas to fuel a central burst of star formation, building up rotational support. The compact and intense star formation ($\HatoUV$>1) ionises this fresh gas, which we are probing with \Ha.}
    \label{fig:HaUV-vs}
\end{figure}

We now investigate how the evolution of sizes at different wavelengths is related to the kinematic evolution of galaxies, namely their ionized gas rotational support $\rotsupp$ and intrinsic velocity dispersion $\disp$. As described in Sec. \ref{sec:morph-kin-mod}, these kinematic quantities are obtained through kinematic modelling with \geko\ and presented in \citet{Danhaive:2025aa}.

In Fig. \ref{fig:HaUV-vs}, we study how the ratio \HatoNUV evolves with the rotational support of the ionized gas, $\rotsupp$. Specifically, we want to relate the rotational support to the movement of galaxies around the main sequence. $\deltaMS$ traces whether galaxies are above or below the star-forming main sequence, on a 10 Myr timescale, but does not contain information on their direction, i.e., $d(\rm SFR)/dt$. Although this is a difficult quantity to constrain, we can approximate it by measuring the ratio of the SFR averaged over the last 10 Myr and over the last 30 Myr ($\rm SFR_{10}/SFR_{30}$). If a galaxy is moving up the main sequence, then its star-formation history (SFH) should be rising, hence  $\rm SFR_{10}/SFR_{30}>1$. Conversely, if it is moving down, then we should observe $\rm SFR_{10}/SFR_{30}<1$. We extract these two quantities from our SED modelling (Sec. \ref{sec:sed-mod}) and colour-code galaxies on Fig. \ref{fig:HaUV-vs} based on whether they have rising (purple) or falling (yellow) SFHs. We note that this cut splits our samples roughly in half, with $55\%$ of galaxies having rising SFHs, which is expected since galaxies should not be found preferentially in either state.

Interestingly, we find that \HatoNUV only correlates with $\rotsupp$ for galaxies with rising SFHs ($\rho = 0.31$, $p<0.001$), with no trend measured for galaxies with falling SFHs ($\rho = 0.07$, $p=0.50$). This can also be seen through the running medians, which are rising and constant, respectively. If we adopt the picture painted in Fig. \ref{fig:HaUV-deltaMS}, where galaxies do not grow inside-out but instead through central starbursts, then Fig. \ref{fig:HaUV-vs} can be interpreted as accreting gas being illuminated by the LyC ionizing photons emitted from the central regions. As the \Ha\ sizes are boosted, they reach further out and could be probing the gas that is fuelling the ongoing star formation. The increase of $\rotsupp$ would then suggest this gas is not infalling radially but rather following circular motions. When galaxies start moving back down towards and below the MS, it implies that star formation is being shut off in the centre, most likely due to outflows and feedback \citep{Tacchella:2016ab,McClymont:2025aa,McClymont:2025ab}. On longer timescales, these processes introduce turbulence in the gas \citep{Mai:2024aa,Danhaive:2025aa} and drive the decrease of $\rotsupp$. Also, with the decrease of the size of the \Ha\ region, we become more sensitive to the central regions of the galaxy where gas is not following circular motions. We discuss this further in Sec. \ref{sec:discussion}.

\subsection{How well does morphology trace kinematics?}\label{sec:morph-disks}

 \begin{figure}
    \centering
    \includegraphics[width=1\linewidth]{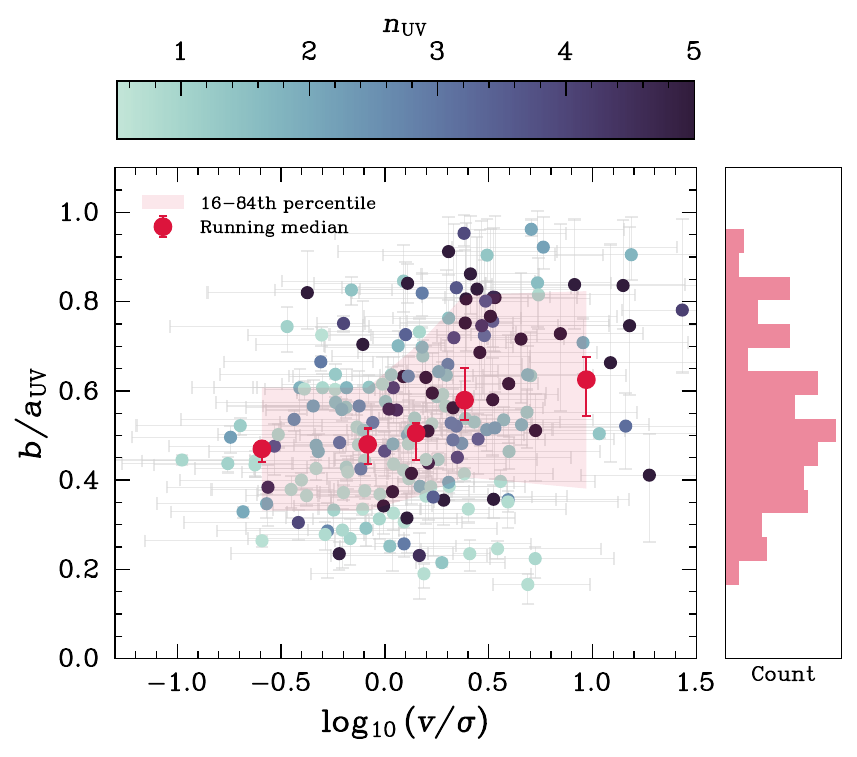}
    \caption{Relation between the rotational support $\vsigma$ and the axis ratio $b/a$ measured from the imaging in the UV. THe points are color-coded based on the measured \sersic index in the UV. We find that $\approx 44\%$ of the systems with $\rm b/a \approx 0$ are not rotationally supported systems, which could point to a population of flattened/prolate systems at high redshift. This is plausible given the non-uniform distribution of the measured axis ratios, suggesting a contribution from spheroids and morphologically prolate systems. }
    \label{fig:ba-vs}
\end{figure}

A key question when measuring gas kinematics, and even more so ionized gas kinematics, is about what exactly are we probing and measuring? For example, mergers can often imitate the kinematic signatures of rotating disks \citep{Robertson:2008aa,Simons:2019aa}, and the rotational support can vary by orders of magnitude depending on the gas tracer used for the measurement \citep{Kohandel:2020aa, Rizzo:2024aa}. At lower spatial resolutions, the effects of outflows and inflows can also impact kinematic measurements, making it difficult to distinguish the origins of the velocities measured. 

In Sec. \ref{sec:kins}, we argue that the trends observed in Fig. \ref{fig:HaUV-vs} could imply that galaxies build up rotational support as they accrete gas to fuel star formation. As the \Ha\ sizes decrease after the starburst, the rotational support decreases due to the smaller scale probed and the turbulence introduced by processes such as feedback from star formation. In this view, the rotational support at high redshift, as measured through \Ha, is more of a probe of the baryon cycle than the build-up of disks. Because of the observational bias promoting high SFRs at low masses in our sample (Fig. \ref{fig:mass-z}), such an interpretation could explain the unexpected decreasing trend observed between $\rotsupp$ and $\logMstar$ in \citet{Danhaive:2025aa}.

Many of the statistical claims of disks in the early universe are based on morphological studies \citep{Ferreira:2022aa,Robertson:2023aa}, which often use projected axis ratios $b/a$ and \sersic indices to infer the disk-like nature of galaxies in the absence of kinematic measurements. As such, galaxies with small values of $b/a$ and/or \sersic indices close to one are often categorized as disks. \citet{Nelson:2023ab} and \citet{Gibson:2024aa} discuss the potential disk-like nature of red ultra-flattened objects (UFOs) uncovered by \textit{JWST}, and further raise the question of how well the morphology traces the underlying kinematics. 

To investigate this further, we plot our measured \Ha\ axis ratios as a function of the rotational support in Fig. \ref{fig:ba-vs}. Across values of $b/a$, we find a diversity of \sersic indices $n$. If the elongated systems in our sample, with $b/a<0.5$, were predominantly disks, we would expect them to have higher values of $\rotsupp$ than those with $b/a>0.5$, which are more spheroidal in nature. However, we instead see the large scatter of values of $\rotsupp$ at fixed $b/a$. This scatter is in part expected as some objects with rotational support could have high values of $b/a$ simply due to being observed face-on. The running medians and the percentiles shown on Fig. \ref{fig:ba-vs} are consistent with a flat trend. 

We find that not all elongated systems are disk-like in nature, but instead that $44\%$ of $b/a<0.5$ systems are flattened systems, which are not rotationally supported. We also plot the distribution of b/a values on Fig. \ref{fig:ba-vs}, and find that it is not uniform but instead peaking around $\rm b/a \approx 0.4-0.5$. Whereas for a population of disks we would expect a flat distribution, such a peaked shape implies the contribution of spheroids and morphologically prolate systems \citep[see e.g.,][]{Robertson:2023aa,Pandya:2024aa}. This result is in agreement with studies such as \citet{Vega-Ferrero:2024aa}, who use the \textsc{tng50} simulations to show that visual and/or morphological classification of galaxies overestimates the fraction of disk-like systems, mislabelling systems which actually have low angular moment and non-oblate structures. \citet{Pandya:2024aa} also discuss the difficulty in constraining 3D shapes from 2D projections, which can lead to misclassifications. More importantly, these elongated non-rotating systems could be the result of filamentary accretion along the cosmic web, with the baryons tracing prolate DM distributions \citep{Ceverino:2015aa}. We also note that although these systems are not rotationally supported, some of them have resolved rotation which is simply secondary with respect to the pressure support. These systems could also be rotating around their minor axis ("kinematically" prolate), although this has never been observed.

The lack of low $\rotsupp$ systems with high $b/a$ is most likely due to the observational limitation of only measuring projected velocities, due to galaxy inclination. The true rotational velocity is obtained by de-projecting the observed velocity, $v_{\rm rot} = v_{\rm obs}/\sin i$, so low values of $i$ (i.e. high values of $b/a$) would boost $v_{\rm rot}$. This results in large uncertainties, as can be seen in Fig. \ref{fig:ba-vs}, but can explain the observed bias.

\section{Discussion} \label{sec:discussion}

Studies with \textit{JWST} have probed the rest-frame UV and optical SMR out to $z\approx 10$, finding a slope of $\alpha \approx 0.2$ ($r\propto M_{\star}^{\alpha}$), which appears to be constant with redshift at $z\approx 0-8$, suggesting a non-evolving relation between stellar and halo mass \citep{Miller:2024aa,Allen:2025aa, Yang:2025aa} for stellar masses $\logMstar \approx 8-11$. In Sec. \ref{sec:res-SMR}, we complement this picture with the SMRs for both stellar continuum (FUV, NUV, and optical) and the ionized gas (through the \Ha\ emission). This offers an unprecedented insight into the growth of galaxies through the baryon cycle, where inflows of gas fuel star formation until feedback kicks in, ejecting the remaining gas. In Sec. \ref{sec:growth-and-kins}, we explore this interplay between gas and stars through the study of size ratios, mainly $\HatoNUV$, and their evolution with star formation and kinematics. 

In this section, we explore the implications of our findings in the greater context of galaxy evolution. In Sec. \ref{sec:disc-SMR}, we interpret our results for the SMR in the context of models of galaxy formation. In Sec. \ref{sec:disc-sizes}, we offer an explanation of the trends found in this paper as the evolution of galaxies around the main sequence through bursts of star formation, and discuss caveats to consider in Sec. \ref{sec:caveats}. 
 
\subsection{The physical drivers of the SMR} \label{sec:disc-SMR}

Analytical theoretical models of galaxy formation provide predictions of the SMR, along with its redshift dependence and intrinsic scatter. In Sec. \ref{sec:slope-evol}, we show that our measurements of the slope $\alpha$ of the SMR are consistent with those found down to $z\approx 0$. As discussed in \citet{Shen:2003aa}, the slope of SMR can be linked to the ratio of stellar to halo mass, and how it depends on the halo mass. Assuming a simple disk model \citep{Fall:1980aa,Mo:1998aa}, a slope of $\alpha = 0.2$ implies that this ratio is not constant, but rather follows a power-law. In fact, assuming $R \propto M_{\rm halo}^{1/3}$ and $R\propto \Mstar^{1/5}$, then $M_{\star}/M_{\rm halo} \propto M_{\rm halo}^{\theta}$ where $\theta = 2/3$.  The lack of redshift evolution in the slope directly implies that this stellar-to-halo mass relation holds out to redshifts of at least $z\approx 6$, beyond which sample sizes decrease, leading to large uncertainties (see Fig. \ref{fig:SMR-Ha-z-slope}). 

However, it is important to consider the stellar mass range probed. \citet{Mowla:2019aa} find that the slope of the SMR is actually mass dependent. In fact, when fitting the SMR with a double power-law, they find that the pivot mass evolves from $\logMstar = 10.5$ to $\logMstar = 11$ from $z=2.5$ to $z=0$. Specifically, the evolution of these pivot masses seem to trace the pivot masses of the SMHM relation from \citet{Behroozi:2019aa}. In our work, we do not probe high enough masses to probe this double power-law behaviour at $z\sim5$, but it is an interesting scope for future study. 

In terms of the redshift dependence, we show in Sec. \ref{sec:z-evol} that the measured flat redshift dependence $\gamma$ found this work, as well the evolution of the median size at $\logMstar = 9.5$ between $z\approx 5$ and $z\approx 0$, suggest that $\gamma$ changes with cosmic time. Specifically, we find evidence for a flattening in the growth of sizes at high redshift, although larger samples, spanning larger redshift ranges, are needed to explore this further. Such a flattening could be attributed to the lack of stable gaseous disks at high redshift \citep{Danhaive:2025aa}, driven by bursty star formation, mergers, and high gas fractions, amongst others. In fact, as galaxies are able to settle into thick disks at cosmic noon and then thin disks at $z<1$, their sizes at fixed mass would increase as the radial extent of the disk increases. On the other hand, at high redshift, galaxies are not able to grow smoothly inside-out due to central starbursts. Overall, a change of $\gamma$ with redshift would imply that different underlying mechanisms influence the sizes of galaxies at different redshifts. Nonetheless, studies of sizes across cosmic time are consistent with the growth of galaxy sizes being driven primarily by the growth of the virial radius of the underlying halo.

Turning to the intrinsic scatter around the SMR, cosmological simulations and theoretical models provide a good framework for understanding its origin. In fact, this scatter $\sigma_{\rm int}$ is often linked to the distribution of spin parameters of dark matter haloes at fixed mass \citep{Mo:1998aa, van-der-Wel:2014ab}. It has been measured within simulations to be around $\sigma_{\rm int} = 0.20-0.25$ dex \citep[e.g.,][]{Burkert:2016aa}, which is consistent with our best-fit values. We find that the offset of galaxies from the SMR follows a log-normal distribution (see Fig. \ref{fig:hist_deltasmr}), which is also consistent with this interpretation. This suggests that, at least at the masses probed by our fiducial fits ($\logMstar>9$), the scatter around the SMR can be attributed to the variation in halo spins, although detailed analysis is required to test this hypothesis \citep[see also][]{Danovich:2012aa,Danovich:2015aa}.

Additionally, the bursty nature of star formation at high redshift could increase the scatter of sizes at fixed mass, particularly at wavelengths more sensitive to recent star formation (i.e., FUV and NUV). Central bursts of star formation are not necessarily spatially correlated to the size or mass of the galaxy, and would have a bigger impact for smaller galaxies, which could help explain the increase in scatter when including galaxies with $\logMstar<9$. The \Ha\ region of ionized gas is predicted to extend beyond the direct star-forming regions (see Sec. \ref{sec:insideout}), and instead is more dependent on the galaxy's SFR which describes the amount of ionizing photons produced. \citet{Simmonds:2025aa} conduct a detailed study of the SFMS and find that the scatter related to short-term burstiness, the difference between the intrinsic scatter of the SFMS between a look-back time of 10 and 50 Myr, is of the order of $\sigma_{\rm ST} \approx 0.2-0.3$ dex in our mass and redshift range. This scatter is consistent with the intrinsic scatter around the SMR, suggesting a potential link between the distribution of sizes at fixed mass and burstiness of star formation.

\subsection{What drives galaxy sizes at $z\approx 4-6$?}\label{sec:disc-sizes}

\begin{figure*}
    \centering
    \includegraphics[width=1\linewidth]{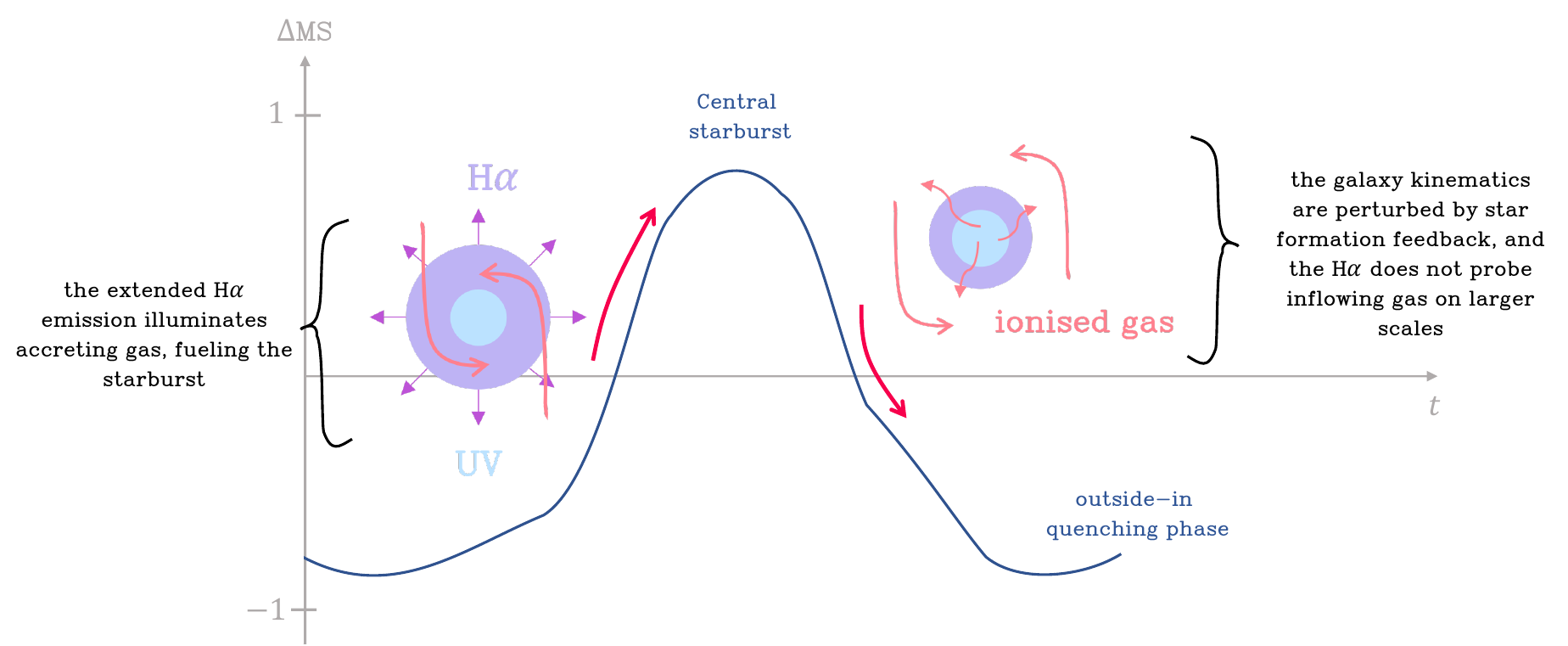}
    \caption{Schematic view of the interplay between \Ha\ sizes and kinematics as proposed in Sec. \ref{sec:discussion}. The large \Ha\ sizes, when compared to the UV continuum tracing young stars, are driven by intense star formation leading up to and during bursts. As the ionizing photons illuminate inflowing gas through \Ha\ emission, we are able to probe the galaxy's rotation as it accretes towards the central regions of the galaxy. Following the burst, the galaxy enters a so-called lulling phase where star formation is quenched in the centre due to stellar feedback. This feedback pushes out the remaining gas and disrupts the kinematics probed by \Ha. Also, as the star formation decreases, so does the radius of the \Ha\ emission, making our observations more sensitive to the turbulent central regions. These oscillations are expected to take place on scales of $\sim 30-100$ Myr \citep{Simmonds:2025aa, McClymont:2025aa}. }
    \label{fig:cartoon}
\end{figure*}

In Secs. \ref{sec:res-SMR}, \ref{sec:insideout}, and \ref{sec:growth-and-kins}, we study the relation between sizes at different rest-frame wavelengths and tracers of ongoing star formation and ionised gas kinematics. Specifically, we have focused on how these different properties vary as galaxies move along the MS ridge-line, and what that can teach us about galaxy growth at high redshift. We discuss here a coherent interpretation of our results, although larger samples sizes and joint studies with simulations are needed to further test its validity. In the local Universe, and up to cosmic noon, weak inside-out growth has been observed through the increase of \Ha\ equivalent width or sSFR with radius \citep{Tacchella:2015aa,Nelson:2016wo,Belfiore:2018aa,Morselli:2019aa}, implying that new stars are forming preferentially in extended star-forming disks. This stable growth is typically observed in galaxies with $\logMstar>10$. At high redshift $z>4$, galaxies have been shown to have disturbed kinematics \citep{Danhaive:2025aa}, bursty SFHs \citep[e.g.][]{Looser:2024aa,Trussler:2024aa,Baker:2025aa,Witten:2025aa}, and high gas fractions \citep{Tacconi:2018aa,Tacconi:2020aa,Parlanti:2023ab,de-Graaff:2024aa, de-Graaff:2024ab}. As such, we do not expect them to grow in the same way as more massive galaxies at lower redshifts. It is important to note that some galaxies at high redshift have thick disks \citep[$10-40\%$;][]{Danhaive:2025aa}, but their low values of $\rotsupp\sim 1-3$ raise the question of whether these disks are long-lived, or whether they are periodically destroyed.

As shown in the right panel of Fig. \ref{fig:HaUV-deltaMS}, we find that the UV traces the optical at $z\approx 4-6$, similar to what has been found in other works at high redshift \citep{Ormerod:2024aa,Ono:2024aa,Yang:2025aa,Allen:2025aa}. We also show in Fig. \ref{fig:dust-mass} that this is unlikely to be caused by central dust obscuration of the UV continuum. Together, these results do not favour inside-out growth. Furthermore, inside-out growth is understood as the formation of a central bulge, followed by the development of an extended star-forming disk. However, as discussed in Sec. \ref{sec:morph-disks}, it is unlikely that star-forming galaxies are forming long-lived disks at $z>4$ in the mass range probed in this work ($\logMstar \approx 7-10$). 

Instead, our work suggests an evolution driven by the movement of galaxies about the SFMS ridge-line. In Fig. \ref{fig:cartoon}, we show a schematic view of the evolution of \Ha\ and UV sizes as galaxies go through a burst of star formation. As galaxies accrete gas towards the central regions, it forms stars. Their LyC photon emission ionizes the gas, forming an \Ha\ halo which expands as the star formation increases. The increase of $\rotsupp$ measured as $\HatoNUV$ increases above the main sequence suggests that the illuminated gas is following circular motions as it is accreting. Once stellar feedback starts to kick in, it injects turbulence into the gas and drives it from the central regions through outflows. These effects could explain the lack of trend as galaxies move down the main sequence (Fig. \ref{fig:HaUV-vs}), as the gas undergoes less ordered motions. Furthermore, depending on the mass of the galaxies and the amount of star formation, the size of the \Ha\ region could be too small to probe the inflow of gas on large scales, which instead could maintain rotational support. Within this framework, \Ha\ acts as a tracer of the baryon cycle within galaxies taking place in the central regions. Cold gas (i.e., CO or [CII]), on the other hand, typically probes larger scales which are less affected by short-term variations in star formation.

Another plausible scenario is that outflows are driving the extended \Ha\ sizes. In fact, studies of [CII] emission with \textit{the Atacama Large Millimeter Array (ALMA)} also find larger [CII] sizes compared to UV, consistent with gas on large scales being illuminated by young stars at the centres of galaxies \citep{Carniani:2018aa,Fujimoto:2020aa}. These [CII] detections could be tracing similar gas to the \Ha. However, this large-scale gas is typically associated to outflowing, and not inflowing, gas, based on the elevated velocity dispersions. Identifying and characterizing outflows at high-redshift is challenging, particularly if they have low velocities, but we expect strong central star formation to result in outflows \citep{Carniani:2024aa} which expel the gas from the central regions \citep{McClymont:2025aa,McClymont:2025ab}. In this work, we favour the inflow scenario mainly based on the fact that the trend of increasing \Ha\ size with $\rotsupp$ is only seen for galaxies with rising SFHs. In the outflow-driven case (Fig. \ref{fig:HaUV-vs}), we would expect the outflows to predominantly kick in when the SFR starts to decline. Observing a trend for galaxies with falling SFHs could have therefore supported the outflow scenario. Nonetheless, outflows could still be contributing to the observed trends and high resolution IFU data for large samples is needed to distinguish these two scenarios. 

\subsection{Caveats} \label{sec:caveats}

There are some caveats of this work that are important to mention in the context of this discussion. As shown in Fig. \ref{fig:mass-z}, we suffer from a bias towards highly star-forming systems at low masses. This is the case for most observational studies, and implies a correlation between $\Mstar$ and $\deltaMS$. We run a partial correlation coefficients (PCC) analysis for the ratio $\HatoNUV$ and find that neither quantity ($\Mstar$ or \deltaMS) can fully account for the other's influence on this ratio. However, we lack the appropriate sample sizes needed for PCC analysis, as can be seen by the bias in the random variable. Aside from the PCC analysis, where we expect sizes to trace stellar mass through the SMR, we do not expect the ratio of sizes to be as sensitive to stellar mass, but rather to physical processes like star formation.

As discussed in Sec. \ref{sec:size-evol}, single \sersic fits could be too simplistic, and hence introduce some difficult-to-quantify biases. This is, for example, the case for galaxies with a bulge and disk, which typically have different \sersic indices for each component. Although $\Reff$ should capture sizes to first order, studies of clumpiness and the use of multi-component modelling are needed to test the validity of single \sersic fits.

There are also uncertainties relating to the kinematic measurements, which could impact our conclusions. They are discussed in detail in \cite{Danhaive:2025aa}, but the main ones are caused by resolution effects and the lack of full 3D data. It is difficult, given grism data, to constrain non-circular motions such as inflows and outflows of gas, which we know will be playing a role in galaxies at high redshift \citep{Carniani:2024aa}. Quantifying their potential imprint on grism data requires careful forward modelling from cosmological simulations, which is beyond the scope of this work. 

Finally, as highlighted on Fig. \ref{fig:mass-z}, our sample's selection suffers from strong degeneracies between stellar mass and offset from the main sequence. Although we attempt to separate the effects of these two quantities in this work, larger samples would be useful to statistically quantify their relative importance further.

\section{Summary \& Conclusions }\label{sec:conclusions}

This work presents a comprehensive morphological analysis of 213 \Ha\ emitters at $z\approx 4-6$ from the FRESCO and CONGRESS \textit{JWST} grism surveys in the GOODS fields. We measure the \Ha\ sizes through morpho-kinematic modelling of the grism data with \geko\, and measure the sizes of the stellar continuum in the FUV, NUV, and optical with \sersic modelling of the JADES NIRCam imaging with \textsc{Pysersic}. We present the SMRs for these different wavelengths and tracers, and discuss their evolution with redshift. We fold in kinematic measurements to discuss the evolution of size ratios with the offset from the star-forming main sequence. We summarize our main findings here:

\begin{itemize}
    \item At $z\approx 5$, we measure the average size of $\logMstar = 9.5$ galaxies and find that the \Ha\ sizes are larger than the stellar continuum, with $\rHa = 1.17 \pm 0.05$ kpc and $r_{\rm cont} \approx 0.9$ kpc, respectively. We find no significant difference between the FUV, NUV, and optical sizes, which are consistent within the uncertainties. 

    \item We find a positive slope for the SMR in the FUV, NUV, and optical of $\alpha \approx 0.2$, consistent with other studies from the local Universe to $z\approx 7$. The slope for the \Ha\ sizes is flatter, $\alpha = 0.15\pm 0.05$, although consistent within the uncertainties. The intrinsic scatter around the SMR is also higher for the stellar continuum, $\sigma_{\rm int}\approx 0.2$, and its value is consistent with the distribution of DM halo spin parameters at constant halo mass. 
    
    \item We do not find evidence for central dust attenuation in our sample, as probed by $\FUVtoNUV$, which implies the positive slope of the SMR is intrinsic and not caused by dust attenuation, as suggested by most large-volume cosmological simulations. Our observational measurements are therefore able to directly place constraints on galaxy formation models.
    
    \item We find no significant evidence for a redshift evolution of sizes, at $\logMstar = 9.5$, within our sample. We cannot draw a firm conclusion due to the limited redshift range of our sample. Within the context of studies at lower redshift, we do however find that sizes increase with cosmic time at fixed stellar mass, as expected. Our measured sizes cannot be reconciled with those reported at $z<2.5$ using a simple power law dependence on the scale factor $(1+z)$ or the Hubble parameter $H(z)$. This points towards a non-constant slope of the evolution of sizes with redshift, with a flattening at $z>3$. 
    
    \item The ratio of \Ha\ to UV sizes increases with $\deltaMS$ above the main sequence, reaching $\log(\HatoNUV) \approx 0.3$ at $\deltaMS>1$. We find this can be caused by the increase of \Ha\ sizes due to increased star formation above the main sequence rather than inside-out growth. In fact, we find no significant evolution of $\FUVtoopt$, which should also trace the extent of young stellar populations compared to older ones. The increase of \Ha\ sizes is consistent with the growth of the Strömgren sphere of LyC photons due to the increase of ionizing radiation during a burst.
    
    \item When dividing galaxies between rising and falling SFHs, $\rm SFR_{10}>SFR_{30}$ and $\rm SFR_{10}<SFR_{30}$, respectively, we find that $\HatoNUV$ has a positive correlation with the rotational support of $\Ha$, $\rotsupp$ for galaxies with rising SFHs ($\rho = 0.31, p < 0.001$). For galaxies with falling SFHs, we find no correlation ($\rho = 0.07, p < 0.50$). We interpret this as the accreting gas, fuelling on-going star formation, being probed by the large \Ha\ radius.

    \item When comparing the observed axis ratio $b/a$ to the rotational support $\rotsupp$, we find that $44\%$ of the systems with $b/a < 0.5$ are not rotationally supported. This suggests that they are not always edge-on disks, but rather a population of galaxies with flattened/prolate morphologies. 

    \end{itemize}

Our work has highlighted the powerful insights that can be drawn from the joint analysis of stellar continuum and \Ha\ sizes, especially when combined with kinematic information. Larger samples are needed for more robust analyses of correlations and for better completeness. Importantly, the forward-modelling of NIRCam imaging and grism data in zoom-in hydrodynamical simulations is crucial to interpret our observed trends and understand the observational biases. We also probe a very bursty regime of galaxy evolution, where kinematics and morphology change on short timescales, making the study of galaxy growth at high redshift more difficult to constrain.

\section*{Acknowledgements}

We thank Hannah Übler for the insightful discussions, Pierluigi Rinaldi and William Baker for the helpful comments, and Christopher Lovell, William Roper, and Xuejian (Jacob) Shen for facilitating access to \textsc{flares} and \textsc{thesan} data. ALD thanks the University of Cambridge Harding Distinguished Postgraduate Scholars Programme and Technology Facilities Council (STFC) Center for Doctoral Training (CDT) in Data intensive science at the University of Cambridge (STFC grant number 2742605) for a PhD studentship. ALD and ST acknowledge support by the Royal Society Research Grant G125142. WM thanks the Science and Technology Facilities Council (STFC) Center for Doctoral Training (CDT) in Data Intensive Science at the University of Cambridge (STFC grant number 2742968) for a PhD studentship. BER acknowledges support from the NIRCam Science Team contract to the University of Arizona, NAS5-02105, and \textit{JWST} Program 3215. S.C acknowledges support by European Union’s HE ERC Starting Grant No. 101040227 - WINGS. CC acknowledges support from the \textit{JWST}/NIRCam Science Team contract to the University of Arizona, NAS5-02105, and \textit{JWST} Program 3215. EE, ZJ, BDJ, MR, CNAW, and YZ acknowlege support from the \textit{JWST}/NIRCam contract to the University of Arizona NAS5-02105. AJB acknowledge funding from the "FirstGalaxies" Advanced Grant from the European Research Council (ERC) under the European Union’s Horizon 2020 research and innovation programme (Grant agreement No. 789056).ECL acknowledges support of an STFC Webb Fellowship (ST/W001438/1). DJE is supported as a Simons Investigator and by \textit{JWST}/NIRCam contract to the University of Arizona, NAS5-02105.  Support for program \#3215 was provided by NASA through a grant from the Space Telescope Science Institute, which is operated by the Association of Universities for Research in Astronomy, Inc., under NASA contract NAS 5-03127. NCV acknowledges support from the Charles and Julia Henry Fund through the Henry Fellowship.

This work is based on observations made with the NASA/ESA Hubble Space Telescope and NASA/ESA/CSA James Webb Space Telescope. The data were obtained from the Mikulski Archive for Space Telescopes at the Space Telescope Science Institute, which is operated by the Association of Universities for Research in Astronomy, Inc., under NASA contract NAS 5-03127 for \textit{JWST}. These observations are associated with program \#1180, 1181, 1210 (JADES), \#1895 (FRESCO), \# 1963 (JEMS) and \#3577 (CONGRESS).
Support for program \#3577 was provided by NASA through a grant from the Space Telescope Science Institute, which is operated by the Association of Universities for Research in Astronomy, Inc., under NASA contract NAS 5-03127. The authors acknowledge the FRESCO team for developing their observing program with a zero-exclusive-access period. The authors acknowledge use of the lux supercomputer at UC Santa Cruz, funded by NSF MRI grant AST 1828315.





\bibliographystyle{mnras}
\bibliography{main} 




\appendix

\section{Additional diagnostics for the SMR}

\begin{figure*}
    \centering
    \includegraphics[width=1\linewidth]{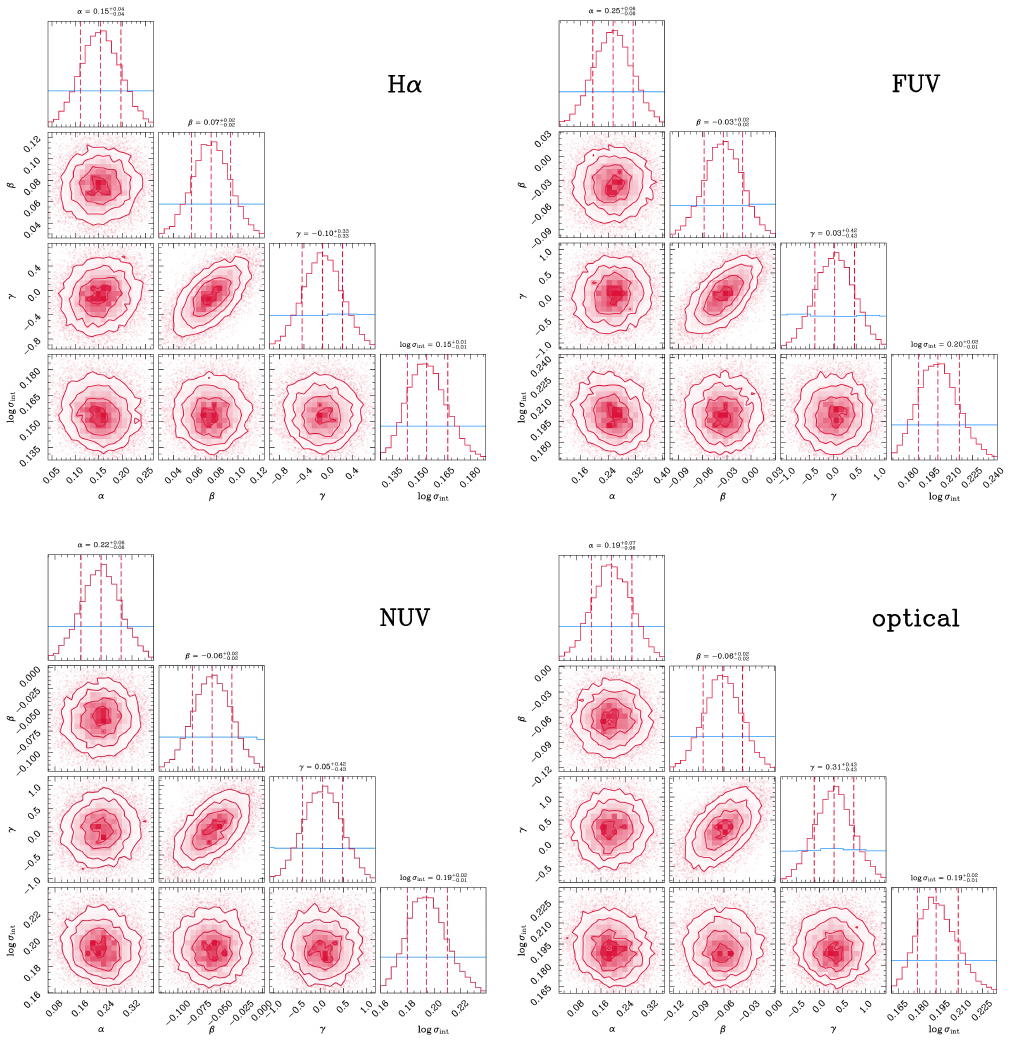}
    \caption{Posterior distributions (red) for the parameters of the SMR (Eq. \ref{eq:SMR}) for the \Ha\, FUV, NUV, and optical sizes, when fitting galaxies with $\logMstar>9$. All of the parameters have uniform priors (blue). }
    \label{fig:post-smr}
\end{figure*}

\begin{figure}
    \centering
    \includegraphics[width=1\linewidth]{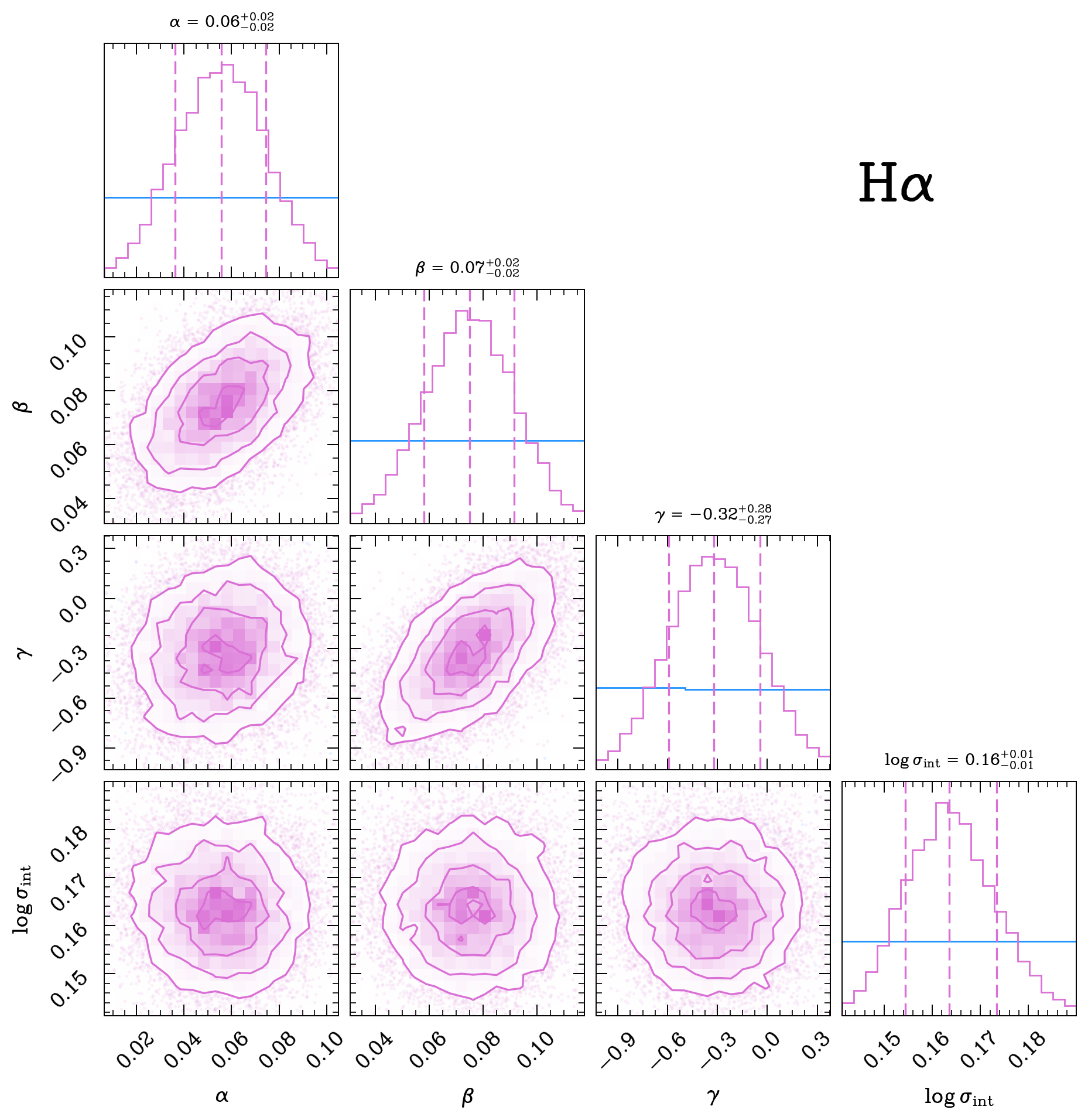}
    \caption{Posterior (purple) and prior (blue) distributions for the parameters of the SMR (Eq. \ref{eq:SMR} for the \Ha\ sizes, when fitting the full sample. }
    \label{fig:post-all}
\end{figure}

\begin{figure}
    \centering
    \includegraphics[width=1\linewidth]{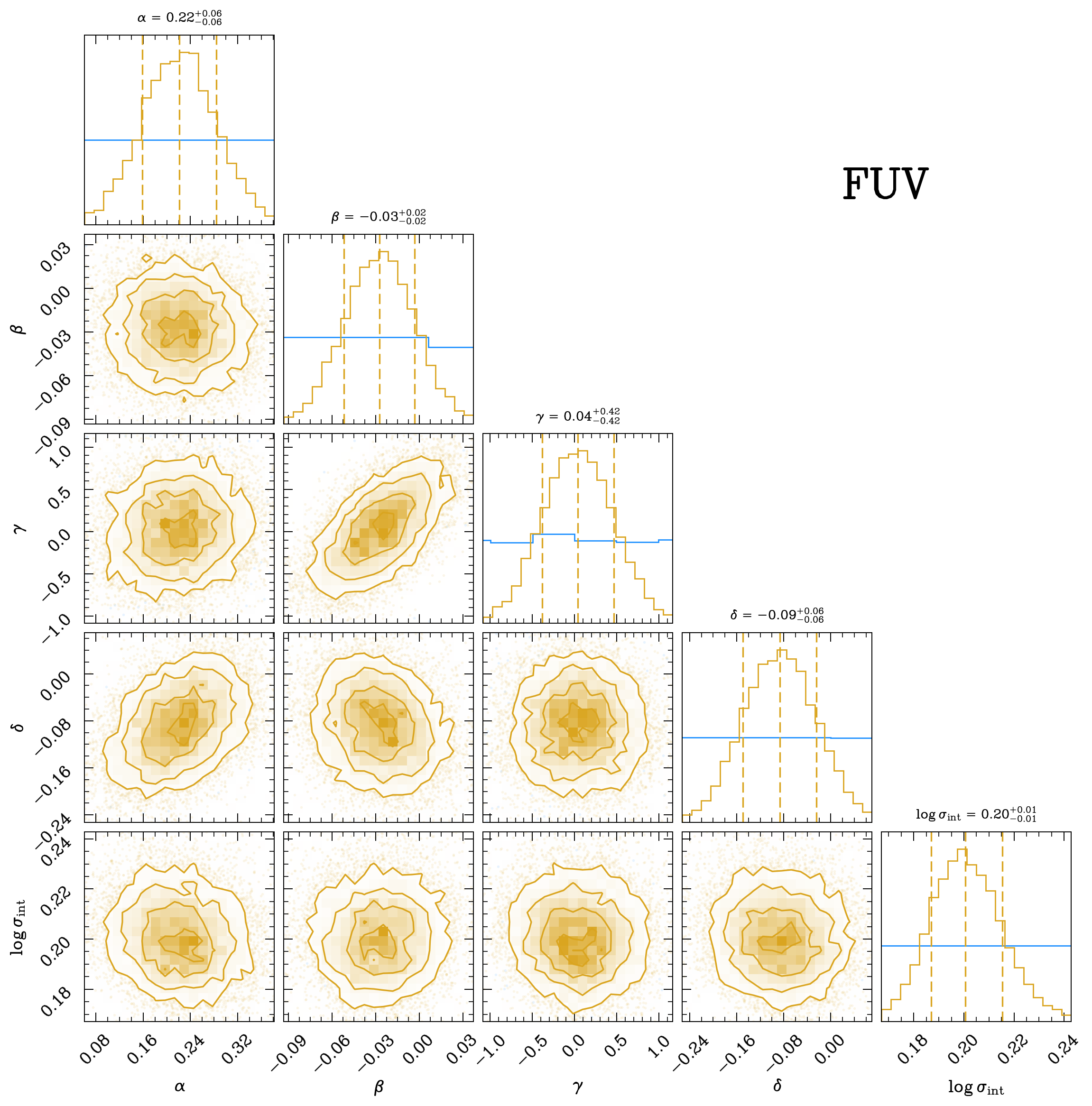}
    \caption{Posterior distributions (red) for the parameters of the SMR (Eq. \ref{eq:SMR} for the  FUV sizes, when including a dependence on $\deltaMS$, which is parametrized through $\delta$. We fit galaxies with $\logMstar>9$. All of the parameters have uniform priors (blue). }
    \label{fig:post-deltams}
\end{figure}

In this section, we present additional diagnostic plots which give insights into the multi-wavelength SMR presented in this works. Fig. \ref{fig:post-smr} shows the posterior distributions of the free parameters of the SMR (Eq. \ref{eq:SMR}) for the \Ha\, FUV, NUV, and optical fiducial (\logMstar>9) fits. All of the parameters are well constrained except the redshift dependence $\gamma$, which suffers from large uncertainties due to the limited redshift range of our sample. In Fig. \ref{fig:post-all} shows the posteriors for the full sample fit of the \Ha\ SMR. As discussed in Sec. \ref{sec:res-SMR}, our measured slope $\alpha$ decreases significantly when we include the low-mass galaxies ($\logMstar<9$) in our fit. On Fig. \ref{fig:post-deltams}, we show posteriors for the FUV SMR but this time we include an additional dependence on the offset of galaxies from the main sequence ($\Delta \rm MS$), parametrized by $\delta$. Despite the large uncertainties, we find evidence for a negative correlation between FUV sizes and $\deltaMS$ ($\delta = -0.09 \pm 0.06$). This suggests a possible compaction of galaxy sizes as they go through bursts of star formation.

\begin{figure}
    \centering
    \includegraphics[width=1\linewidth]{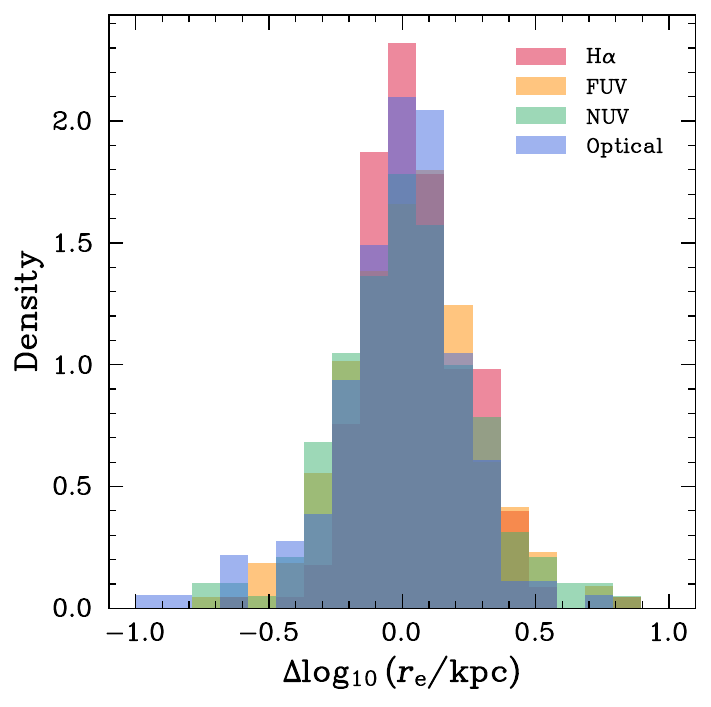}
    \caption{Distribution of the offset of galaxies from the SMR, $\Delta \log_{10}(\Reff) = \log_{10}(\Reff) - \log_{10}(\rm SMR)$, for \Ha, FUV, NUV, and optical sizes. We find that the scatter follows a log-normal distribution.}
    \label{fig:hist_deltasmr}
\end{figure}

As shown by the posterior distribution, we fit for the intrinsic scatter $\sigma_{\rm int}$ around the SMRs. To study the shape of offset of galaxies from the SMR, $\Delta \log (\Reff/\rm kpc)$, we plot a histogram of its value for our full sample on Fig. \ref{fig:hist_deltasmr}. It is clear that at all wavelengths probed, the scatter around the SMR follows a log-normal distribution.

\begin{figure}
    \centering
    \includegraphics[width=1\linewidth]{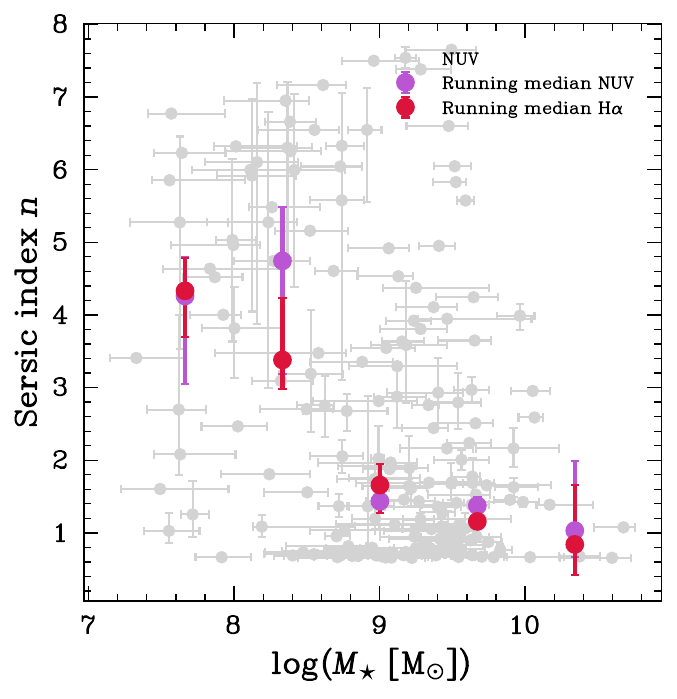}
    \caption{\sersic index as a function of stellar mass for the NUV morphologies derived from our \geko\ fits. We also include the running medians for the \Ha, which show the same trends. }
    \label{fig:n_ha}
\end{figure}

The definition of \sersic profiles (Eq. \ref{eq:Sersic_profile}) implies a form of degeneracy between the \sersic index $n$ and the effective radius $\Reff$. In order to explore this bias as a function of stellar mass, we plot the \Ha\ and NUV \sersic indices for our sample on Fig. \ref{fig:n_ha}. We find a large scatter across all stellar masses, but find that the medians converge to $n\sim1$ at high masses $\logMstar>9$. On the other hand, the lower-mass galaxies have typically larger \sersic indices ($n\sim4$), which implies a steeper central light profile. This could bias size measurements at low masses to higher values.

\bsp	
\label{lastpage}
\end{document}